\documentclass[sigconf]{acmart}

\usepackage{pifont}
\usepackage{soul}
\usepackage{subcaption}
\usepackage{siunitx}

\newcommand{\cmark}{\ding{51}}%
\newcommand{\xmark}{\ding{55}}%

\usepackage{pgfplots, pgfplotstable}
\usepackage{tikz}
\pgfplotsset{compat=newest}
\pgfplotsset{plot coordinates/math parser=false}
\newlength\fheight
\newlength\fwidth
\usetikzlibrary{plotmarks, shapes, patterns, decorations.pathreplacing, backgrounds,calc, arrows,arrows.meta,spy, matrix,external}
\usepackage{tikzscale}
\usepackage{multicol}
\usepackage{multirow}

\usepackage[edges]{forest}
\ExplSyntaxOn
\cs_generate_variant:Nn \str_case:nnF { V }
\cs_new_eq:NN \strCase \str_case:VnF
\ExplSyntaxOff

\definecolor{folderbg}{RGB}{124,166,198}
\definecolor{folderborder}{RGB}{110,144,169}
\newlength\Size
\tikzset{%
  folder/.pic={%
    \filldraw [draw=folderborder, top color=folderbg!50, bottom color=folderbg] (-1.05*\Size,0.2\Size+5pt) rectangle ++(.75*\Size,-0.2\Size-5pt);
    \filldraw [draw=folderborder, top color=folderbg!50, bottom color=folderbg] (-1.15*\Size,-\Size) rectangle (1.15*\Size,\Size);
  },
  file/.pic={%
    \filldraw [draw=folderborder, top color=folderbg!5, bottom color=folderbg!10] (-\Size,.4*\Size+5pt) coordinate (a) |- (\Size,-1.2*\Size) coordinate (b) -- ++(0,1.6*\Size) coordinate (c) -- ++(-5pt,5pt) coordinate (d) -- cycle (d) |- (c) ;
  },
}
\forestset{%
  declare autowrapped toks={pic me}{},
  pic dir tree/.style={%
    for tree={%
      folder,
      font=\ttfamily,
      grow'=0,
    },
    before typesetting nodes={%
      for tree={%
        edge label+/.option={pic me},
      },
    },
  },
  pic me set/.code n args=2{%
    \forestset{%
      #1/.style={%
        inner xsep=2\Size,
        pic me={pic {#2}},
      }
    }
  },
  pic me set={directory}{folder},
  pic me set={file}{file},
}

\acmYear{2024}\copyrightyear{2024}
\acmConference[MMSys '24]{ACM Multimedia Systems Conference 2024}{April 15--18, 2024}{Bari, Italy}
\acmBooktitle{ACM Multimedia Systems Conference 2024 (MMSys '24), April 15--18, 2024, Bari, Italy}
\acmDOI{10.1145/3625468.3652190}
\acmISBN{979-8-4007-0412-3/24/04}

\usepackage{epsfig}
\usepackage[capitalize]{cleveref}

\usepackage{glossaries}
\newacronym{fps}{FPS}{frames per second}
\newacronym{dl}{DL}{Downlink}
\newacronym{gnb}{gNB}{Next Generation Node Base}
\newacronym{hmd}{HMD}{Head-Mounted Display}
\newacronym{iot}{IoT}{Internet of Things}
\newacronym{pcap}{PCAP}{Packet CAPture}
\newacronym{qoe}{QoE}{Quality of Experience}
\newacronym{qos}{QoS}{Quality of Service}
\newacronym{ssq}{SSQ}{Simulator Sickness Questionnaire}
\newacronym{usb}{USB}{Universal Serial Bus}
\newacronym{ul}{UL}{Uplink}
\newacronym{xr}{XR}{eXtended Reality}
\newacronym{vr}{VR}{Virtual Reality}
\newacronym{dt}{DT}{Digital Twin}
\newacronym{bpda}{BPDA}{Boston Planning and Development Administration}
\newacronym{gis}{GIS}{Geographic Information System}
\newacronym{lod}{LOD}{Level of Detail}
\newacronym{dtm}{DTM}{Digital Terrain Model}
\newacronym{crs}{CRS}{Coordinate Reference System}
\newacronym{wgs84}{WGS84}{World Geodetic System 1984}
\newacronym{csv}{CSV}{Comma Separated Values}
\newacronym{nad83}{NAD 83}{North American Datum 1983}
\newacronym{nadv88}{NADV 88}{North American Vertical Datum 1988}
\newacronym{iso}{ISO}{International Organization for Standardization}
\newacronym{ad}{AD}{Autonomous Driving}
\newacronym{rt}{RT}{Ray Tracer}
\newacronym{em}{EM}{Electromagnetic}
\newacronym{sumo}{SUMO}{Simulation of Urban MObility}
\newacronym{v2x}{V2X}{Vehicle-to-Everything}

\newacronym{3gpp}{3GPP}{3rd Generation Partnership Project}
\newacronym{4g}{4G}{4th generation}
\newacronym{5g}{5G}{5th generation}
\newacronym{6g}{6G}{6th generation}
\newacronym{5gc}{5GC}{5G Core}
\newacronym{adc}{ADC}{Analog to Digital Converter}
\newacronym{aerpaw}{AERPAW}{Aerial Experimentation and Research Platform for Advanced Wireless}
\newacronym{ai}{AI}{Artificial Intelligence}
\newacronym{aimd}{AIMD}{Additive Increase Multiplicative Decrease}
\newacronym{am}{AM}{Acknowledged Mode}
\newacronym{amc}{AMC}{Adaptive Modulation and Coding}
\newacronym{amf}{AMF}{Access and Mobility Management Function}
\newacronym{aops}{AOPS}{Adaptive Order Prediction Scheduling}
\newacronym{api}{API}{Application Programming Interface}
\newacronym{apn}{APN}{Access Point Name}
\newacronym{ap}{AP}{Application Protocol}
\newacronym{aqm}{AQM}{Active Queue Management}
\newacronym{ausf}{AUSF}{Authentication Server Function}
\newacronym{avc}{AVC}{Advanced Video Coding}
\newacronym{awgn}{AGWN}{Additive White Gaussian Noise}
\newacronym{balia}{BALIA}{Balanced Link Adaptation Algorithm}
\newacronym{bbu}{BBU}{Base Band Unit}
\newacronym{bdp}{BDP}{Bandwidth-Delay Product}
\newacronym{ber}{BER}{Bit Error Rate}
\newacronym{bf}{BF}{Beamforming}
\newacronym{bler}{BLER}{Block Error Rate}
\newacronym{brr}{BRR}{Bayesian Ridge Regressor}
\newacronym{bs}{BS}{Base Station}
\newacronym{bsr}{BSR}{Buffer Status Report}
\newacronym{bss}{BSS}{Business Support System}
\newacronym{ca}{CA}{Carrier Aggregation}
\newacronym{caas}{CaaS}{Connectivity-as-a-Service}
\newacronym{cb}{CB}{Code Block}
\newacronym{cc}{CC}{Congestion Control}
\newacronym{ccid}{CCID}{Congestion Control ID}
\newacronym{cco}{CC}{Carrier Component}
\newacronym{cdd}{CDD}{Cyclic Delay Diversity}
\newacronym{cdf}{CDF}{Cumulative Distribution Function}
\newacronym{cdn}{CDN}{Content Distribution Network}
\newacronym{cli}{CLI}{Command-line Interface}
\newacronym{cn}{CN}{Core Network}
\newacronym{codel}{CoDel}{Controlled Delay Management}
\newacronym{comac}{COMAC}{Converged Multi-Access and Core}
\newacronym{cord}{CORD}{Central Office Re-architected as a Datacenter}
\newacronym{cornet}{CORNET}{COgnitive Radio NETwork}
\newacronym{cosmos}{COSMOS}{Cloud Enhanced Open Software Defined Mobile Wireless Testbed for City-Scale Deployment}
\newacronym{cots}{COTS}{Commercial Off-the-Shelf}
\newacronym{cp}{CP}{Control Plane}
\newacronym{cyp}{CP}{Cyclic Prefix}
\newacronym{up}{UP}{User Plane}
\newacronym{cpu}{CPU}{Central Processing Unit}
\newacronym{cqi}{CQI}{Channel Quality Information}
\newacronym{cr}{CR}{Cognitive Radio}
\newacronym{cran}{CRAN}{Cloud \gls{ran}}
\newacronym{csi}{CSI}{Channel State Information}
\newacronym{csirs}{CSI-RS}{Channel State Information - Reference Signal}
\newacronym{cu}{CU}{Central Unit}
\newacronym{d2tcp}{D$^2$TCP}{Deadline-aware Data center TCP}
\newacronym{d3}{D$^3$}{Deadline-Driven Delivery}
\newacronym{dac}{DAC}{Digital to Analog Converter}
\newacronym{dag}{DAG}{Directed Acyclic Graph}
\newacronym{das}{DAS}{Distributed Antenna System}
\newacronym{dash}{DASH}{Dynamic Adaptive Streaming over HTTP}
\newacronym{dc}{DC}{Dual Connectivity}
\newacronym{dccp}{DCCP}{Datagram Congestion Control Protocol}
\newacronym{dce}{DCE}{Direct Code Execution}
\newacronym{dci}{DCI}{Downlink Control Information}
\newacronym{dctcp}{DCTCP}{Data Center TCP}
\newacronym{dmr}{DMR}{Deadline Miss Ratio}
\newacronym{dmrs}{DMRS}{DeModulation Reference Signal}
\newacronym{drlcc}{DRL-CC}{Deep Reinforcement Learning Congestion Control}
\newacronym{drs}{DRS}{Discovery Reference Signal}
\newacronym{du}{DU}{Distributed Unit}
\newacronym{e2e}{E2E}{end-to-end}
\newacronym{earfcn}{EARFCN}{E-UTRA Absolute Radio Frequency Channel Number}
\newacronym{ecaas}{ECaaS}{Edge-Cloud-as-a-Service}
\newacronym{ecn}{ECN}{Explicit Congestion Notification}
\newacronym{edf}{EDF}{Earliest Deadline First}
\newacronym{embb}{eMBB}{Enhanced Mobile Broadband}
\newacronym{empower}{EMPOWER}{EMpowering transatlantic PlatfOrms for advanced WirEless Research}
\newacronym{enb}{eNB}{evolved Node Base}
\newacronym{endc}{EN-DC}{E-UTRAN-\gls{nr} \gls{dc}}
\newacronym{epc}{EPC}{Evolved Packet Core}
\newacronym{eps}{EPS}{Evolved Packet System}
\newacronym{es}{ES}{Edge Server}
\newacronym{etsi}{ETSI}{European Telecommunications Standards Institute}
\newacronym[firstplural=Estimated Times of Arrival (ETAs)]{eta}{ETA}{Estimated Time of Arrival}
\newacronym{eutran}{E-UTRAN}{Evolved Universal Terrestrial Access Network}
\newacronym{faas}{FaaS}{Function-as-a-Service}
\newacronym{fapi}{FAPI}{Functional Application Platform Interface}
\newacronym{fdd}{FDD}{Frequency Division Duplexing}
\newacronym{fdm}{FDM}{Frequency Division Multiplexing}
\newacronym{fdma}{FDMA}{Frequency Division Multiple Access}
\newacronym{fed4fire}{FED4FIRE+}{Federation 4 Future Internet Research and Experimentation Plus}
\newacronym{fir}{FIR}{Finite Impulse Response}
\newacronym{fit}{FIT}{Future \acrlong{iot}}
\newacronym{fpga}{FPGA}{Field Programmable Gate Array}
\newacronym{fr2}{FR2}{Frequency Range 2}
\newacronym{fs}{FS}{Fast Switching}
\newacronym{fscc}{FSCC}{Flow Sharing Congestion Control}
\newacronym{ftp}{FTP}{File Transfer Protocol}
\newacronym{fw}{FW}{Flow Window}
\newacronym{ge}{GE}{Gaussian Elimination}
\newacronym{gop}{GOP}{Group of Pictures}
\newacronym{gpr}{GPR}{Gaussian Process Regressor}
\newacronym{gpu}{GPU}{Graphics Processing Unit}
\newacronym{gtp}{GTP}{GPRS Tunneling Protocol}
\newacronym{gtpc}{GTP-C}{GPRS Tunnelling Protocol Control Plane}
\newacronym{gtpu}{GTP-U}{GPRS Tunnelling Protocol User Plane}
\newacronym{gtpv2c}{GTPv2-C}{\gls{gtp} v2 - Control}
\newacronym{gw}{GW}{Gateway}
\newacronym{harq}{HARQ}{Hybrid Automatic Repeat reQuest}
\newacronym{hetnet}{HetNet}{Heterogeneous Network}
\newacronym{hh}{HH}{Hard Handover}
\newacronym{hol}{HOL}{Head-of-Line}
\newacronym{hqf}{HQF}{Highest-quality-first}
\newacronym{hss}{HSS}{Home Subscription Server}
\newacronym{http}{HTTP}{HyperText Transfer Protocol}
\newacronym{ia}{IA}{Initial Access}
\newacronym{iab}{IAB}{Integrated Access and Backhaul}
\newacronym{ic}{IC}{Incident Command}
\newacronym{ietf}{IETF}{Internet Engineering Task Force}
\newacronym{imsi}{IMSI}{International Mobile Subscriber Identity}
\newacronym{imt}{IMT}{International Mobile Telecommunication}
\newacronym{ip}{IP}{Internet Protocol}
\newacronym{itu}{ITU}{International Telecommunication Union}
\newacronym{kpi}{KPI}{Key Performance Indicator}
\newacronym{kpm}{KPM}{Key Performance Measurement}
\newacronym{kvm}{KVM}{Kernel-based Virtual Machine}
\newacronym{los}{LOS}{Line-of-Sight}
\newacronym{lsm}{LSM}{Link-to-System Mapping}
\newacronym{lstm}{LSTM}{Long Short Term Memory}
\newacronym{lte}{LTE}{Long Term Evolution}
\newacronym{lxc}{LXC}{Linux Container}
\newacronym{m2m}{M2M}{Machine to Machine}
\newacronym{mac}{MAC}{Medium Access Control}
\newacronym{manet}{MANET}{Mobile Ad Hoc Network}
\newacronym{mano}{MANO}{Management and Orchestration}
\newacronym{mc}{MC}{Multi-Connectivity}
\newacronym{mcc}{MCC}{Mobile Cloud Computing}
\newacronym{mchem}{MCHEM}{Massive Channel Emulator}
\newacronym{mcs}{MCS}{Modulation and Coding Scheme}
\newacronym{mec}{MEC}{Multi-access Edge Computing}
\newacronym{mec2}{MEC}{Mobile Edge Cloud}
\newacronym{mfc}{MFC}{Mobile Fog Computing}
\newacronym{mgen}{MGEN}{Multi-Generator}
\newacronym{mi}{MI}{Mutual Information}
\newacronym{mib}{MIB}{Master Information Block}
\newacronym{miesm}{MIESM}{Mutual Information Based Effective SINR}
\newacronym{mimo}{MIMO}{Multiple Input, Multiple Output}
\newacronym{ml}{ML}{Machine Learning}
\newacronym{mlr}{MLR}{Maximum-local-rate}
\newacronym[plural=\gls{mme}s,firstplural=Mobility Management Entities (MMEs)]{mme}{MME}{Mobility Management Entity}
\newacronym{mmtc}{mMTC}{Massive Machine-Type Communications}
\newacronym{mmwave}{mmWave}{millimeter wave}
\newacronym{mpdccp}{MP-DCCP}{Multipath Datagram Congestion Control Protocol}
\newacronym{mptcp}{MPTCP}{Multipath TCP}
\newacronym{mr}{MR}{Maximum Rate}
\newacronym{mrdc}{MR-DC}{Multi \gls{rat} \gls{dc}}
\newacronym{mse}{MSE}{Mean Square Error}
\newacronym{mss}{MSS}{Maximum Segment Size}
\newacronym{mt}{MT}{Mobile Termination}
\newacronym{mtd}{MTD}{Machine-Type Device}
\newacronym{mtu}{MTU}{Maximum Transmission Unit}
\newacronym{mumimo}{MU-MIMO}{Multi-user \gls{mimo}}
\newacronym{mvno}{MVNO}{Mobile Virtual Network Operator}
\newacronym{nalu}{NALU}{Network Abstraction Layer Unit}
\newacronym{nas}{NAS}{Network Attached Storage}
\newacronym{nat}{NAT}{Network Address Translation}
\newacronym{nbiot}{NB-IoT}{Narrow Band IoT}
\newacronym{nfv}{NFV}{Network Function Virtualization}
\newacronym{nfvi}{NFVI}{Network Function Virtualization Infrastructure}
\newacronym{ni}{NI}{Network Interfaces}
\newacronym{nic}{NIC}{Network Interface Card}
\newacronym{nlos}{NLOS}{Non-Line-of-Sight}
\newacronym{now}{NOW}{Non Overlapping Window}
\newacronym{nsm}{NSM}{Network Service Mesh}
\newacronym{nrf}{NRF}{Network Repository Function}
\newacronym{nsa}{NSA}{Non Stand Alone}
\newacronym{nse}{NSE}{Network Slicing Engine}
\newacronym{nssf}{NSSF}{Network Slice Selection Function}
\newacronym{o2i}{O2I}{Outdoor to Indoor}
\newacronym{oai}{OAI}{OpenAirInterface}
\newacronym{oaicn}{OAI-CN}{\gls{oai} \acrlong{cn}}
\newacronym{oairan}{OAI-RAN}{\acrlong{oai} \acrlong{ran}}
\newacronym{oam}{OAM}{Operations, Administration and Maintenance}
\newacronym{ofdm}{OFDM}{Orthogonal Frequency Division Multiplexing}
\newacronym{olia}{OLIA}{Opportunistic Linked Increase Algorithm}
\newacronym{omec}{OMEC}{Open Mobile Evolved Core}
\newacronym{onap}{ONAP}{Open Network Automation Platform}
\newacronym{onf}{ONF}{Open Networking Foundation}
\newacronym{onos}{ONOS}{Open Networking Operating System}
\newacronym{oom}{OOM}{\gls{onap} Operations Manager}
\newacronym{opnfv}{OPNFV}{Open Platform for \gls{nfv}}
\newacronym{orbit}{ORBIT}{Open-Access Research Testbed for Next-Generation Wireless Networks}
\newacronym{os}{OS}{Operating System}
\newacronym{oss}{OSS}{Operations Support System}
\newacronym{pa}{PA}{Position-aware}
\newacronym{pase}{PASE}{Prioritization, Arbitration, and Self-adjusting Endpoints}
\newacronym{pawr}{PAWR}{Platforms for Advanced Wireless Research}
\newacronym{pbch}{PBCH}{Physical Broadcast Channel}
\newacronym{pcef}{PCEF}{Policy and Charging Enforcement Function}
\newacronym{pcfich}{PCFICH}{Physical Control Format Indicator Channel}
\newacronym{pcrf}{PCRF}{Policy and Charging Rules Function}
\newacronym{pdcch}{PDCCH}{Physical Downlink Control Channel}
\newacronym{pdcp}{PDCP}{Packet Data Convergence Protocol}
\newacronym{pdsch}{PDSCH}{Physical Downlink Shared Channel}
\newacronym{pdu}{PDU}{Packet Data Unit}
\newacronym{pf}{PF}{Proportional Fair}
\newacronym{pgw}{PGW}{Packet Gateway}
\newacronym{phich}{PHICH}{Physical Hybrid ARQ Indicator Channel}
\newacronym{phy}{PHY}{Physical}
\newacronym{pmch}{PMCH}{Physical Multicast Channel}
\newacronym{pmi}{PMI}{Precoding Matrix Indicators}
\newacronym{powder}{POWDER}{Platform for Open Wireless Data-driven Experimental Research}
\newacronym{ppo}{PPO}{Proximal Policy Optimization}
\newacronym{ppp}{PPP}{Poisson Point Process}
\newacronym{prach}{PRACH}{Physical Random Access Channel}
\newacronym{prb}{PRB}{Physical Resource Block}
\newacronym{psnr}{PSNR}{Peak Signal to Noise Ratio}
\newacronym{pss}{PSS}{Primary Synchronization Signal}
\newacronym{pucch}{PUCCH}{Physical Uplink Control Channel}
\newacronym{pusch}{PUSCH}{Physical Uplink Shared Channel}
\newacronym{qam}{QAM}{Quadrature Amplitude Modulation}
\newacronym{qci}{QCI}{\gls{qos} Class Identifier}
\newacronym{quic}{QUIC}{Quick UDP Internet Connections}
\newacronym{rach}{RACH}{Random Access Channel}
\newacronym[firstplural=Radio Access Technologies (RATs)]{rat}{RAT}{Radio Access Technology}
\newacronym{rbg}{RBG}{Resource Block Group}
\newacronym{rcn}{RCN}{Research Coordination Network}
\newacronym{rc}{RC}{RAN Control}
\newacronym{rec}{REC}{Radio Edge Cloud}
\newacronym{red}{RED}{Random Early Detection}
\newacronym{renew}{RENEW}{Reconfigurable Eco-system for Next-generation End-to-end Wireless}
\newacronym{rf}{RF}{Radio Frequency}
\newacronym{rfc}{RFC}{Request for Comments}
\newacronym{rfr}{RFR}{Random Forest Regressor}
\newacronym{ric}{RIC}{\gls{ran} Intelligent Controller}
\newacronym{rlc}{RLC}{Radio Link Control}
\newacronym{rlf}{RLF}{Radio Link Failure}
\newacronym{rlnc}{RLNC}{Random Linear Network Coding}
\newacronym{rmr}{RMR}{RIC Message Router}
\newacronym{rmse}{RMSE}{Root Mean Squared Error}
\newacronym{rnis}{RNIS}{Radio Network Information Service}
\newacronym{rr}{RR}{Round Robin}
\newacronym{rrc}{RRC}{Radio Resource Control}
\newacronym{rrm}{RRM}{Radio Resource Management}
\newacronym{rru}{RRU}{Remote Radio Unit}
\newacronym{rs}{RS}{Remote Server}
\newacronym{rsrp}{RSRP}{Reference Signal Received Power}
\newacronym{rsrq}{RSRQ}{Reference Signal Received Quality}
\newacronym{rss}{RSS}{Received Signal Strength}
\newacronym{rssi}{RSSI}{Received Signal Strength Indicator}
\newacronym{rtt}{RTT}{Round Trip Time}
\newacronym{ru}{RU}{Radio Unit}
\newacronym{rw}{RW}{Receive Window}
\newacronym{rx}{RX}{Receiver}
\newacronym{s1ap}{S1AP}{S1 Application Protocol}
\newacronym{sa}{SA}{standalone}
\newacronym{sack}{SACK}{Selective Acknowledgment}
\newacronym{sap}{SAP}{Service Access Point}
\newacronym{sc2}{SC2}{Spectrum Collaboration Challenge}
\newacronym{scef}{SCEF}{Service Capability Exposure Function}
\newacronym{sch}{SCH}{Secondary Cell Handover}
\newacronym{scoot}{SCOOT}{Split Cycle Offset Optimization Technique}
\newacronym{sctp}{SCTP}{Stream Control Transmission Protocol}
\newacronym{sdap}{SDAP}{Service Data Adaptation Protocol}
\newacronym{sdk}{SDK}{Software Development Kit}
\newacronym{sdm}{SDM}{Space Division Multiplexing}
\newacronym{sdma}{SDMA}{Spatial Division Multiple Access}
\newacronym{sdn}{SDN}{Software-defined Networking}
\newacronym{sdr}{SDR}{Software-defined Radio}
\newacronym{seba}{SEBA}{SDN-Enabled Broadband Access}
\newacronym{sgsn}{SGSN}{Serving GPRS Support Node}
\newacronym{sgw}{SGW}{Service Gateway}
\newacronym{si}{SI}{Study Item}
\newacronym{sib}{SIB}{Secondary Information Block}
\newacronym{sinr}{SINR}{Signal to Interference plus Noise Ratio}
\newacronym{sip}{SIP}{Session Initiation Protocol}
\newacronym{siso}{SISO}{Single Input, Single Output}
\newacronym{sla}{SLA}{Service Level Agreement}
\newacronym{sm}{SM}{Service Model}
\newacronym{smf}{SMF}{Session Management Function}
\newacronym{smo}{SMO}{Service Management and Orchestration}
\newacronym{sms}{SMS}{Short Message Service}
\newacronym{smsgmsc}{SMS-GMSC}{\gls{sms}-Gateway}
\newacronym{snr}{SNR}{Signal-to-Noise-Ratio}
\newacronym{son}{SON}{Self-Organizing Network}
\newacronym{sptcp}{SPTCP}{Single Path TCP}
\newacronym{srb}{SRB}{Service Radio Bearer}
\newacronym{srn}{SRN}{Standard Radio Node}
\newacronym{srs}{SRS}{Sounding Reference Signal}
\newacronym{ss}{SS}{Synchronization Signal}
\newacronym{sss}{SSS}{Secondary Synchronization Signal}
\newacronym{st}{ST}{Spanning Tree}
\newacronym{svc}{SVC}{Scalable Video Coding}
\newacronym{tb}{TB}{Transport Block}
\newacronym{tcp}{TCP}{Transmission Control Protocol}
\newacronym{tdd}{TDD}{Time Division Duplexing}
\newacronym{tdm}{TDM}{Time Division Multiplexing}
\newacronym{tdma}{TDMA}{Time Division Multiple Access}
\newacronym{tfl}{TfL}{Transport for London}
\newacronym{tfrc}{TFRC}{TCP-Friendly Rate Control}
\newacronym{tft}{TFT}{Traffic Flow Template}
\newacronym{tgen}{TGEN}{Traffic Generator}
\newacronym{tip}{TIP}{Telecom Infra Project}
\newacronym{tm}{TM}{Transparent Mode}
\newacronym{to}{TO}{Telco Operator}
\newacronym{tr}{TR}{Technical Report}
\newacronym{trp}{TRP}{Transmitter Receiver Pair}
\newacronym{ts}{TS}{Technical Specification}
\newacronym{tti}{TTI}{Transmission Time Interval}
\newacronym{ttt}{TTT}{Time-to-Trigger}
\newacronym{tx}{TX}{Transmitter}
\newacronym{uas}{UAS}{Unmanned Aerial System}
\newacronym{uav}{UAV}{Unmanned Aerial Vehicle}
\newacronym{udm}{UDM}{Unified Data Management}
\newacronym{udp}{UDP}{User Datagram Protocol}
\newacronym{udr}{UDR}{Unified Data Repository}
\newacronym{ue}{UE}{User Equipment}
\newacronym{uhd}{UHD}{\gls{usrp} Hardware Driver}
\newacronym{um}{UM}{Unacknowledged Mode}
\newacronym{uml}{UML}{Unified Modeling Language}
\newacronym{upa}{UPA}{Uniform Planar Array}
\newacronym{upf}{UPF}{User Plane Function}
\newacronym{urllc}{URLLC}{Ultra Reliable and Low Latency Communications}
\newacronym{usa}{U.S.}{United States}
\newacronym{usim}{USIM}{Universal Subscriber Identity Module}
\newacronym{usrp}{USRP}{Universal Software Radio Peripheral}
\newacronym{utc}{UTC}{Urban Traffic Control}
\newacronym{vim}{VIM}{Virtualization Infrastructure Manager}
\newacronym{vm}{VM}{Virtual Machine}
\newacronym{vnf}{VNF}{Virtual Network Function}
\newacronym{volte}{VoLTE}{Voice over \gls{lte}}
\newacronym{voltha}{VOLTHA}{Virtual OLT HArdware Abstraction}
\newacronym{vran}{vRAN}{Virtualized \gls{ran}}
\newacronym{vss}{VSS}{Video Streaming Server}
\newacronym{wbf}{WBF}{Wired Bias Function}
\newacronym{wf}{WF}{Waterfilling}
\newacronym{wg}{WG}{Working Group}
\newacronym{wlan}{WLAN}{Wireless Local Area Network}
\newacronym{osm}{OSM}{Open Source \gls{nfv} Management and Orchestration}
\newacronym{pnf}{PNF}{Physical Network Function}
\newacronym{drl}{DRL}{Deep Reinforcement Learning}
\newacronym{mtc}{MTC}{Machine-type Communications}
\newacronym{mns}{MnS}{Management Services}
\newacronym{ves}{VES}{\gls{vnf} Event Stream}
\newacronym{ei}{EI}{Enrichment Information}
\newacronym{fh}{FH}{Fronthaul}
\newacronym{fft}{FFT}{Fast Fourier Transform}
\newacronym{laa}{LAA}{Licensed-Assisted Access}
\newacronym{plfs}{PLFS}{Physical Layer Frequency Signals}
\newacronym{ptp}{PTP}{Precision Time Protocol}
\newacronym{cbrs}{CBRS}{Citizen Broadband Radio Service}

\newacronym{cif}{CI}{cyberinfrastructure}
\newacronym{sonic}{SONiC}{Software for Open Networking in the Cloud}
\newacronym{ocp}{OCP}{Open Compute Project}
\newacronym{snmp}{SNMP}{Simple Network Management Protocol}
\newacronym{raid}{RAID}{redundant array of independent disks}
\newacronym{nfs}{NFS}{Network File Storage}
\newacronym{ci}{CI}{Continuous Integration}
\newacronym{cd}{CD}{Continuous Deployment}
\newacronym{dtn}{DTN}{Data Transfer Node}

\newacronym{dns}{DNS}{Domain Name Service}
\newacronym{nrpe}{NRPE}{Nagios Remote Plugin Executor}
\newacronym{ldap}{LDAP}{Lightweight Directory Access Protocol}
\newacronym{lan}{LAN}{Local Area Network}
\newacronym{vlan}{VLAN}{Virtual LAN}

\newacronym{ipmi}{IPMI}{Intelligent Platform Management Interface}
\newacronym{tor}{ToR}{Top-of-the-Rack}
\newacronym{lmn}{LMN}{Large Memory Node}
\newacronym{bgp}{BGP}{Border Gateway Protocol}
\newacronym{dhcp}{DHCP}{Dynamic Host Configuration Protocol}
\newacronym{vrf}{VRF}{Virtual Routing and Forwarding}
\newacronym{vpn}{VPN}{Virtual Private Network}
\newacronym{rma}{RMA}{Return Merchandise Authorization}
\newacronym{hpc}{HPC}{High Performance Compute}

\newacronym{nu}{NU}{Northeastern University}
\newacronym{asic}{ASIC}{Application-specific Integrated Circuit}
\newacronym{rdma}{RDMA}{Remote Direct Memory Access}
\newacronym{roce}{RoCE}{RDMA over Converged Ethernet}
\newacronym{ovs}{OVS}{Open vSwitch}
\newacronym{frr}{FRR}{Free Range Routing}
\newacronym{ups}{UPS}{Uninterruptible Power Supply}

\newacronym{ntia}{NTIA}{National Telecommunications and Information Administration}
\newacronym{irb}{IRB}{Institutional Review Board}
\newacronym{doi}{DOI}{Digital Object Identifier}

\newacronym{sdo}{SDO}{Standard-Development Organization}
\newacronym{ose}{OSE}{Open Source Ecosystem}
\newacronym{osc}{OSC}{O-RAN Software Community}
\newacronym{dop}{DOP}{Director of Operations}
\newacronym{pm}{PM}{Program Manager}
\newacronym{excom}{EXCOM}{Executive Committee}
\newacronym{iiot}{IIoT}{Industrial \gls{iot}}
\newacronym{lf}{LF}{Linux Foundation}

\newacronym{wiot}{WIoT}{Institute for the Wireless Internet of Things}

\newacronym{otic}{OTIC}{Open Testing \& Integration Centre}

\newacronym{nofo}{NOFO}{Notice of Funding Opportunity}

\newacronym{onr}{ONR}{Office of Naval Research}
\newacronym{afosr}{AFOSR}{Air Force Office of Scientific Research}
\newacronym{afrl}{AFRL}{Air Force Research Laboratory}
\newacronym{arl}{ARL}{Army Research Laboratory}

\newacronym{arc}{ARC}{Aerial Research Cloud}

\newacronym{mno}{MNO}{Mobile Network Operator}
\newacronym{ct}{CT}{Continuous Testing}
\newacronym{oci}{OCI}{Open Container Initiative}

\newacronym[plural=RANs]{ran}{RAN}{Radio Access Network}
\newacronym{pii}{PII}{Personally Identifiable Information}
\newacronym{cves}{CVEs}{Common Vulnerabilities and Exposures}
\newacronym{n-rt-ric}{Near-RT RIC}{Near-Real-Time RAN Intelligent Controller}
\newacronym{o-cu}{O-CU-CP}{O-RAN Central Unit}
\newacronym{o-cu-cp}{O-CU-CP}{O-RAN Central Unit - Control Plane}
\newacronym{o-cu-up}{O-CU-UP}{O-RAN Central Unit - User Plane}
\newacronym{o-du}{O-DU}{O-RAN Distrubuted Unit}
\newacronym{o-ru}{O-RU}{O-RAN Radio Unit}
\newacronym{oran}{O-RAN}{Open Radio Access Network}
\newacronym{sast}{SAST}{Static application security testing}
\newacronym{rbac}{RBAC}{Role-Based Access Control}

\usepackage{xcolor}
\definecolor{green_cool}{rgb}{0.0, 0.5, 0.0}
\definecolor{blue_cool}{rgb}{0.12, 0.32, 1}
\definecolor{red_cool}{rgb}{0.5, 0.0, 0.0}
\definecolor{light_blue}{rgb}{00, 0.9, 0.9}

\usepackage{xspace}
\newcommand{\projName}{BostonTwin\xspace}

\usepackage{makecell}

\usepackage{soul}

\begin{document}

\title{\projName: the Boston Digital Twin for Ray-Tracing in 6G Networks}

\author{Paolo Testolina}
\email{p.testolina@northeastern.edu}
\affiliation{%
 \institution{Northeastern University}
 \city{Boston}
 \state{MA}
 \country{USA}}

\author{Michele Polese}
\email{m.polese@northeastern.edu}
\affiliation{%
 \institution{Northeastern University}
 \city{Boston}
 \state{MA}
 \country{USA}}
 
\author{Pedram Johari}
\email{p.johari@northeastern.edu}
\affiliation{%
 \institution{Northeastern University}
 \city{Boston}
 \state{MA}
 \country{USA}}
 
\author{Tommaso Melodia}
\email{t.melodia@northeastern.edu}
\affiliation{%
 \institution{Northeastern University}
 \city{Boston}
 \state{MA}
 \country{USA}}
 
\renewcommand{\shortauthors}{Testolina, et al.}

\begin{abstract}
  Digital twins are now a staple of wireless networks design and evolution.
  Creating an accurate digital copy of a real system offers numerous opportunities to study and analyze its performance and issues. It also allows designing and testing new solutions in a risk-free environment, and applying them back to the real system after validation.
  A candidate technology that will heavily rely on digital twins for design and deployment is 6G, which promises robust and ubiquitous networks for \gls{xr} and immersive communications solutions.
  In this paper, we present \projName, a dataset that merges a high-fidelity 3D model of the city of Boston, MA, with the existing geospatial data on cellular base stations deployments, in a ray-tracing-ready format.
  Thus, \projName enables not only the instantaneous rendering and programmatic access to the building models, but it also allows for an accurate representation of the electromagnetic propagation environment in the real-world city of Boston.
  The level of detail and accuracy of this characterization is crucial to designing 6G networks that can support the strict requirements of sensitive and high-bandwidth applications, such as \gls{xr} and immersive communication.
\end{abstract}



\begin{teaserfigure}
  \includegraphics[width=\textwidth]{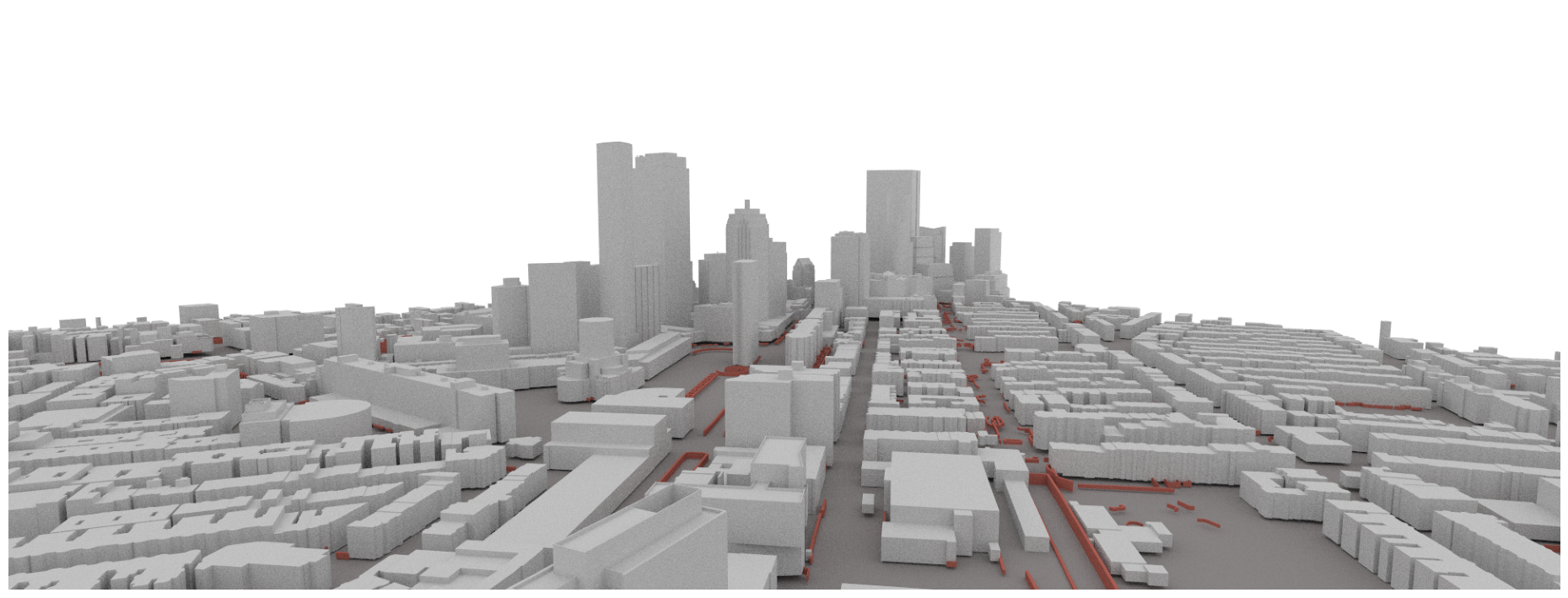}
  \caption{Rendering of Downtown Boston using the \projName open data and code available at~\cite{bostontwin_repo}.}
  \label{fig:teaser}
\end{teaserfigure}


\maketitle

\begin{tikzpicture}[remember picture,overlay]
  \node[anchor=north,yshift=-10pt] at (current page.north) {\parbox{\dimexpr\textwidth-\fboxsep-\fboxrule\relax}{
  \centering\footnotesize This paper has been accepted for presentation at ACM Multimedia Systems Conference 2024 (MMSys '24).
  Please cite it as: P.Testolina, M. Polese, P. Johari, and T. Melodia, "BostonTwin: the Boston Digital Twin for Ray-Tracing in 6G Networks," in Proceedings of the ACM Multimedia Systems Conference 2024, ser. MMSys '24. Bari, Italy: Association for Computing Machinery, 2024.}};
\end{tikzpicture}

\section{Introduction}

The ever-increasing complexity of networks and digital systems calls for new comprehensive methodologies for their design and optimization. In this context, the concept of \glspl{dt} has emerged as a powerful tool and abstraction in a variety of applications, including networking and multimedia solutions~\cite{itu-r-2160,atis2023roadmap, grieves2015dt,Jones2020characteristing}. Twinning recreates real-world scenarios in a digital format with an extremely high level of realism. In this way, the \gls{dt} can be used as a safe playground where to test and evaluate different configurations for the real-world system (i) without impacting its performance and the quality of service to the end-users; and (ii) possibly anticipating and evaluating multiple versions of the real-world system in a faster-than-real-time environment. In addition, a high-fidelity digital replica allows for the collection of data that can be used to train \gls{ai} models, with flexibility and potential for automation which are difficult to achieve in the real world.

\glspl{dt} have been leveraged in prior studies on networking~\cite{he2022resource,zeb2022industrial,zhao2020idt,latif2022dte,nguyen2021dt,zhao2022interlink,dai2022adaptive}, to improve optimization and configuration of wireless networks, planning monitoring solutions, and improve vehicular network and satellite systems performance, and in multimedia systems~\cite{qi2023digital,huang2023digital,zhao2023qoe,aloqaily2023integrating}, to profile quality of service for mobile users streaming videos, manage resources for multicast streaming, and optimize quality of experience in \gls{xr} systems. Implementing digital twins, however, is a challenging task, especially in wireless systems where the digital world needs to accurately replicate real-world propagation, protocol stacks, applications, and user patterns. We took a first step toward defining a high-fidelity \gls{dt} for wireless networks in~\cite{villa2024colosseum}, introducing the Colosseum wireless network emulator as a \gls{dt} of over-the-air wireless networks and toolchains for the twinning of wireless channels and networks.

In this paper, we take a further step in the creation of \glspl{dt} that can realistically replicate wireless scenarios. We introduce \projName, the first fully automated toolchain, and dataset for city-scale twinning, focusing on the city of Boston through open data released by the \gls{bpda}. We implement and open source a set of \glspl{api} for the integration of this 3D model in popular ray tracers, e.g., NVIDIA Sionna~\cite{sionna, hoydis2023sionnart}, and the fusion of the city mesh with additional objects in their real-world locations, e.g., base stations, for cellular networks design. The BostonTwin 3D model allows for capturing fine-grained details---\emph{at a city scale}---that can fundamentally alter the propagation of wireless signals, and thus the accuracy of a \gls{dt}. The code and data are publicly available at~\cite{bostontwin_repo} and~\cite{bostontwin_data}.

Our contributions toward creating \projName are as follows:

\begin{itemize} 
	\item We convert the \glspl{bpda} 3D building model~\cite{3d_data} to a metric-based model in the PLY format widely used by ray tracers, e.g., Mitsuba, which allows the programmatic and fast rendering of 3D scenes for graphical or \gls{rf} ray tracing~\cite{jakob2022mitsuba3}. We provide the code to perform such conversion and also the processed dataset at~\cite{bostontwin_repo}. 

    \item We provide an \gls{api} to interact with and manipulate \projName. The \gls{api} enables the deployment and geolocation of nodes and points of interest, the dynamic tuning of the level of details and accuracy of the \gls{dt} based on the user requirements, and the extraction of sub-scenes from the complete city scenario.

    \item Leveraging the \projName \gls{api}, we extend the city-scale 3D model with information on the location of points of interest for \gls{rf} ray tracing. Notably, we consider the locations of wireless cellular base stations in the city of Boston, also released as open data~\cite{antennas_data}.
    This methodology can be adapted to integrate additional elements, e.g., mobile receivers, other wireless systems, or sources of interference.

    \item The dataset in the PLY format and the \projName \glspl{api} directly feed information on the 3D mesh to the NVIDIA Sionna electromagnetic ray tracer, which we use to profile propagation in different tiles of the city of Boston and support for \gls{6g} wireless multimedia use cases (e.g., \gls{xr}).

\end{itemize}

The \projName approach and \glspl{api} introduce a simplified approach to interact with \gls{dt} models, together with an accurate city-scale 3D dataset.
Compared to existing \glspl{dt}~\cite{bhatia2023efficient,nokia2024dt,ericssson2021intelligent}, \projName supports an unprecedented scale, is open-source, and is integrated with other open-source, state-of-the-art tools, thus supporting open science and research.

%
%
The rest of the paper is organized as follows. \cref{sec:framework} introduces the data used in \projName, as well as the Python classes that can be used to manipulate the 3D models and geolocated points of interest. \cref{sec:workflow} describes the workflow we propose to use and interact with \projName and \gls{rf} ray tracers. \cref{sec:results} discusses a use cases of \projName, i.e., the large-scale evaluation of coverage and data rates for immersive communications in a urban environment. Finally, \cref{sec:conclusions} concludes the paper and discusses future extensions. 

\section{The \projName Framework}
\label{sec:framework}

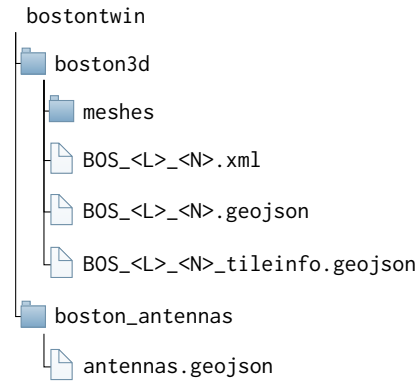
\begin{figure}[t]
\begin{forest}
  pic dir tree,
  where level=0{}{
    file,
  },
  where level=3{}{
    directory,
  }
[{bostontwin}
  [boston3d
        [meshes
        ]
        [BOS\_<L>\_<N>.xml, file]
        [BOS\_<L>\_<N>.geojson, file]
        [BOS\_<L>\_<N>\_tileinfo.geojson, file]
    ]
  [boston\_antennas
        [antennas.geojson, file]
  ]
]
\end{forest}
\caption{Dataset organization.}
\label{fig:dataset_dirs}
\end{figure}

The \projName framework consists of the \gls{api} and of two main data sources, namely, the 3D building model and the base station geographic information.
The corresponding data is accessible through the \texttt{BostonModel} and \texttt{BostonAntennas} Python classes, that provide several utility functions to manage, manipulate, and interact with the dataset. The dataset itself is organized as reported in \cref{fig:dataset_dirs}.
In the following subsections, we report a brief overview of the two classes that support the \projName operations, whereas the \gls{api} and the workflow to interact with the digital twin are described in \cref{sec:workflow}.

\subsection{The \texttt{BostonModel} Class}
\label{ssec:BostonModel}
\begin{table}[t]
    \centering
    \footnotesize
    \caption{Location and 3D model information for the tiles covering the central area of Boston.}
    \label{tab:tiles_info}
    \begin{tabular}{|c|c|c|c|c|}
    \hline
        \textbf{Scene Name} & \textbf{Lon.} & \textbf{Lat.} & \textbf{\# Models} & \textbf{\# Triangles} \\ \hline
        BOS\_F\_4 & 42.3570902827 & -71.1025189571 & 181 & 22113\\ \hline
        BOS\_F\_5 & 42.3433701314 & -71.1026051428 & 1535 & 199456\\ \hline
        BOS\_F\_6 & 42.3296499389 & -71.1026912911 & 3691 & 313610\\ \hline
        BOS\_F\_7 & 42.3159297061 & -71.1027774021 & 4817 & 382937\\ \hline
        BOS\_F\_8 & 42.3022094336 & -71.1028634758 & 1393 & 95710\\ \hline
        BOS\_F\_9 & 42.2884891223 & -71.1029495121 & 1175 & 64484\\ \hline
        BOS\_G\_3 & 42.3707449543 & -71.0839300113 & 2 & 52\\ \hline
        BOS\_G\_4 & 42.3570248589 & -71.0840202471 & 1527 & 130612\\ \hline
        BOS\_G\_5 & 42.3433047216 & -71.0841104437 & 4640 & 393083\\ \hline
        BOS\_G\_6 & 42.3295845431 & -71.0842006013 & 3657 & 238585\\ \hline
        BOS\_G\_7 & 42.3158643242 & -71.0842907198 & 4669 & 333976\\ \hline
        BOS\_G\_8 & 42.3021440657 & -71.0843807992 & 3388 & 230992\\ \hline
        BOS\_G\_9 & 42.2884237683 & -71.0844708396 & 3711 & 229814\\ \hline
        BOS\_H\_3 & 42.3706765403 & -71.0654273276 & 2743 & 240304\\ \hline
        BOS\_H\_4 & 42.3569564595 & -71.0655215761 & 3173 & 314862\\ \hline
        BOS\_H\_5 & 42.3432363368 & -71.0656157838 & 3056 & 263616\\ \hline
        BOS\_H\_6 & 42.3295161729 & -71.0657099505 & 2449 & 132728\\ \hline
        BOS\_H\_7 & 42.3157959687 & -71.0658040765 & 5337 & 398158\\ \hline
        BOS\_H\_8 & 42.3020757248 & -71.0658981616 & 5080 & 342727\\ \hline
        BOS\_H\_9 & 42.288355442 & -71.065992206 & 5346 & 335896\\ \hline
        BOS\_I\_3 & 42.37060515 & -71.0469246849 & 1598 & 177828\\ \hline
        BOS\_I\_4 & 42.3568850846 & -71.0470229461 & 855 & 150213\\ \hline
        BOS\_I\_5 & 42.3431649771 & -71.0471211646 & 1733 & 161432\\ \hline
        BOS\_I\_6 & 42.3294448285 & -71.0472193406 & 4498 & 224807\\ \hline
        BOS\_I\_7 & 42.3157246395 & -71.047317474 & 2253 & 152828\\ \hline
        BOS\_I\_8 & 42.3020044108 & -71.0474155649 & 868 & 60083\\ \hline
        BOS\_I\_9 & 42.2882841434 & -71.0475136132 & 3772 & 215304\\ \hline
        BOS\_J\_3 & 42.3705307836 & -71.0284220849 & 2708 & 162831\\ \hline
        BOS\_J\_5 & 42.3430906425 & -71.0286265881 & 1041 & 89440\\ \hline
        BOS\_J\_6 & 42.3293705098 & -71.0287287732 & 2108 & 114492\\ \hline
        BOS\_J\_7 & 42.3156503367 & -71.028830914 & 82 & 8840\\ \hline
        \textbf{Total} &  &  & 83086 & 6181813 \\ \hline
    \end{tabular}
    
\end{table}

The \texttt{BostonModel} class contains the information and the methods to access the 3D model of the buildings of the city of Boston.
The data was obtained through the publicly available \acrfull{bpda} Citywide 3D Smart Model~\cite{3d_data}, which in turn originated from the 2011 planimetry of Boston, and has been regularly updated since then using data from \gls{bpda} Urban Design models, CyberCity 3D, and LiDAR sources.
Currently, the data is organized and managed according to the citySchema model~\cite{cityschema} by the \gls{bpda} \gls{gis} Lab.
Thanks to the open-source management format, which fulfills the requirements of the ISO Reference Model for Open Archival Systems~\cite{lee2010open}, the data is easily maintained, updated, and expanded.
Furthermore, as more cities adopt the citySchema standard and the 3D models get further refined, the \gls{api} and the corresponding digital twining framework provided through this work can be enhanced and expanded to compatible cities and scenarios.

The 3D data covers the entire city of Boston and is organized in $N$ square tiles of 1524 by 1524~m (5000 by 5000~ft).
The tile are named according to the format BOS\_<L>\_<N>, with L$\in\{\text{A},\text{B},\dots,\text{O}\}$ and N$\in\{1,2,\dots,13\}$.
The complete list of currently available tiles is reported in \cref{tab:tiles_info}.
For each tile, different data formats for the 3D models are available, e.g., OBJ Waveform and Sketchup files, and a Model Catalog in \gls{csv}.
While the building coordinates in the Model Catalog are expressed as (Longitude, Latitude), following the \gls{wgs84}, or EPSG:4326 reference, the OBJ models use a custom \gls{crs}, as reported in the \gls{bpda} documentation.
In particular, the Massachusetts State Plane in feet is employed as the geographical projection plane, with \gls{nad83} and \gls{nadv88} as Horizontal and Vertical Datum, respectively.
The origin of the custom \gls{crs} is then set at (731100, 2902900) feet of the Massachusetts State Plane (\gls{wgs84}: (Longitude: 71.223391 W, Latitude: 42.213379 N)).
\begin{table}[t]
    \centering
    \footnotesize
    \caption{\gls{lod} as defined by the citySchema format.}
    \label{tab:lod}
    \begin{tabular}{|c|c|}
    \hline
    \textbf{CODE} & \textbf{Description}\\
    \hline
        LOD 0 & Polygon Footprint\\
    \hline
        LOD 1 & Extruded Polygon Footprint\\
    \hline
        LOD 1.5 & \makecell{Massing model made\\ from extruded roof prints\\ when a structure with parts\\ that have different heights.}\\
    \hline
        LOD 2 & \makecell{3D roof detail, extruded\\ to the ground along drip-line.}\\
    \hline
        LOD 3 & \makecell{Model portrays undercuts\\ where appropriate.}\\
    \hline
        LOD 3.25 & \makecell{Architectural details indicated\\ by materials or image textures.}\\
    \hline
        LOD 3.5 & \makecell{Building model expresses\\ the location of windows and entryways\\ as 3D indentations.}\\
    \hline
        LOD 4 & \makecell{Model is divided horizontally\\ as individual stories.}\\
    \hline
        LOD 4.5 & \makecell{Model divides interior spaces:\\ rooms or zones.}\\
    \hline
    \end{tabular}

\end{table}
\newcommand{\sizefiggg}{0.23}
\begin{figure}
    \setlength\tabcolsep{0pt}
    \renewcommand{\arraystretch}{0}
    \centering
    \begin{tabular}{c|c|c|c|c|}
      \multicolumn{5}{c}{
\begin{tikzpicture}
\definecolor{lod_0}{rgb}{0.12156862745098039, 0.4666666666666667, 0.7058823529411765}%
\definecolor{lod_1}{rgb}{1.0, 0.4980392156862745, 0.054901960784313725}%
\definecolor{lod_1.5}{rgb}{0.17254901960784313, 0.6274509803921569, 0.17254901960784313}%
\definecolor{lod_2}{rgb}{0.8392156862745098, 0.15294117647058825, 0.1568627450980392}%
\definecolor{lod_3}{rgb}{0.5803921568627451, 0.403921568627451, 0.7411764705882353}%
\definecolor{lod_3.25}{rgb}{0.5490196078431373, 0.33725490196078434, 0.29411764705882354}%
\definecolor{lod_3.5}{rgb}{0.8901960784313725, 0.4666666666666667, 0.7607843137254902}%
\definecolor{lod_4}{rgb}{0.4980392156862745, 0.4980392156862745, 0.4980392156862745}%
\definecolor{lod_4.5}{rgb}{0.7372549019607844, 0.7411764705882353, 0.13333333333333333}%

\begin{axis}[%
width=0,
height=0,
at={(0,0)},
xmin=0,
xmax=0,
xtick={},
ymin=0,
ymax=0,
ytick={},
scale only axis,
axis background/.style={fill=white},
legend style={legend cell align=center, align=center, draw=white!15!black,at={(0,0)},anchor=south west, /tikz/every even column/.append style={column sep = 0.5cm},font=\scriptsize},
legend columns = 5
]

\addlegendimage{scatter, only marks, color=lod_0, line width=1.pt};
\addlegendentry{LOD 0}

\addlegendimage{scatter, only marks, color=lod_1, line width=1.pt};
\addlegendentry{LOD 1}

\addlegendimage{scatter, only marks, color=lod_1.5, line width=1.pt};
\addlegendentry{LOD 1.5}

\addlegendimage{scatter, only marks, color=lod_2, line width=1.pt};
\addlegendentry{LOD 2}

\addlegendimage{scatter, only marks, color=lod_3, line width=1.pt};
\addlegendentry{LOD 3}

\addlegendimage{scatter, only marks, color=lod_3.25, line width=1.pt};
\addlegendentry{LOD 3.25}

\addlegendimage{scatter, only marks, color=lod_3.5, line width=1.pt};
\addlegendentry{LOD 3.5}

\addlegendimage{scatter, only marks, color=lod_4, line width=1.pt};
\addlegendentry{LOD 4}

\addlegendimage{scatter, only marks, color=lod_4.5, line width=1.pt};
\addlegendentry{LOD 4.5}
\end{axis}

\end{tikzpicture}}\\
     & F & G & H & I\\\hline
    \rotatebox{90}{\qquad \ \ \ 3} & & &
    \includegraphics[width=\sizefiggg\linewidth]{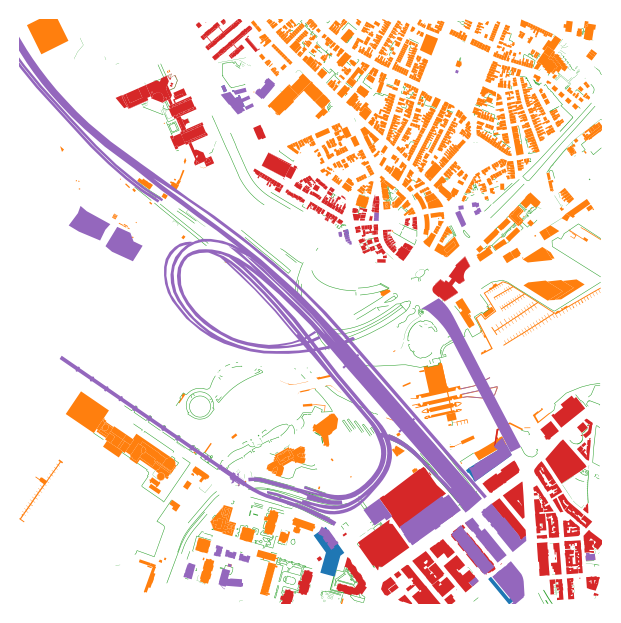} &
    \includegraphics[width=\sizefiggg\linewidth]{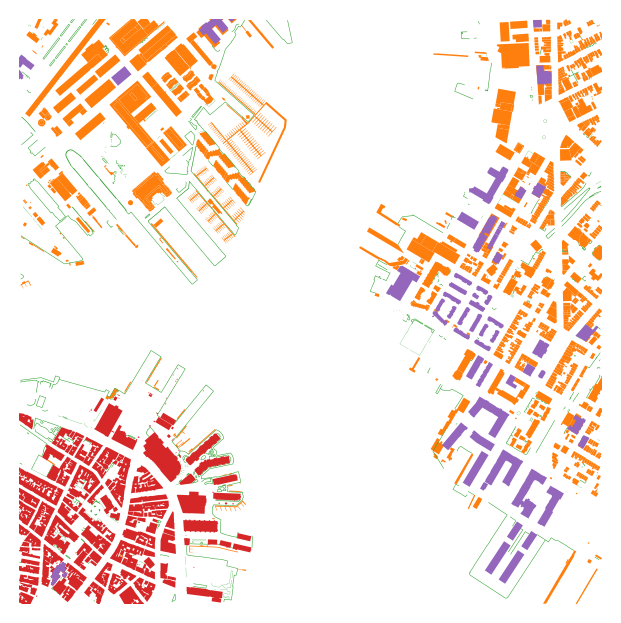}\\\hline
    \rotatebox{90}{\qquad \ \ \ 4} & \includegraphics[width=\sizefiggg\linewidth]{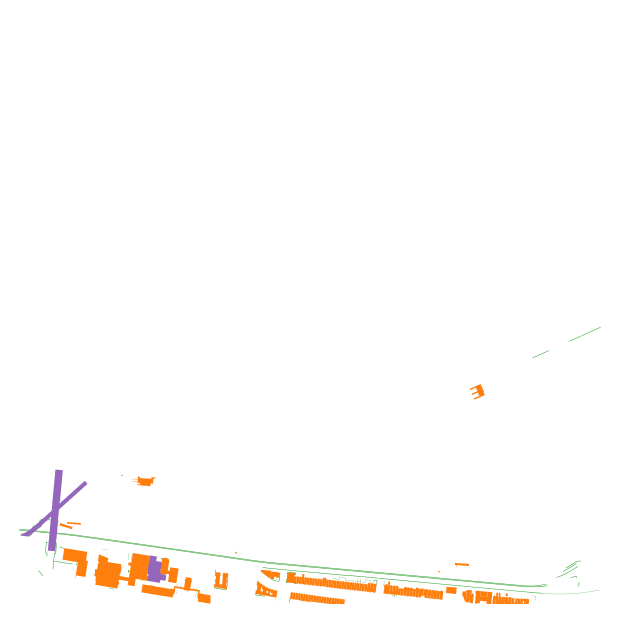} &
    \includegraphics[width=\sizefiggg\linewidth]{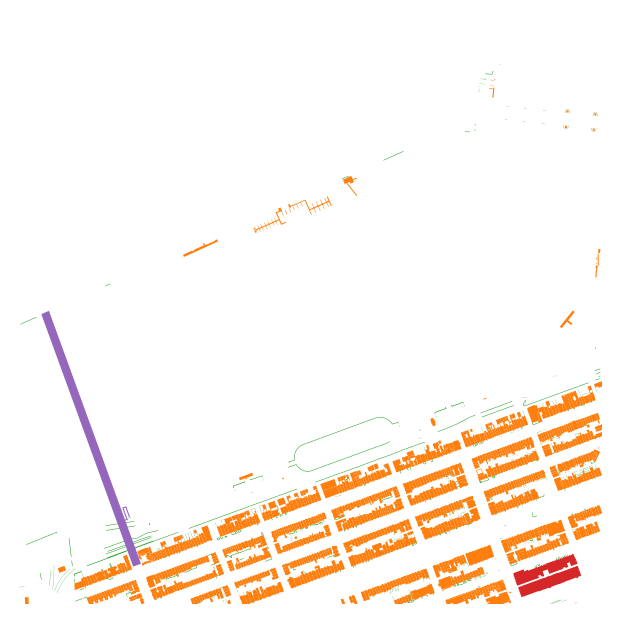} &
    \includegraphics[width=\sizefiggg\linewidth]{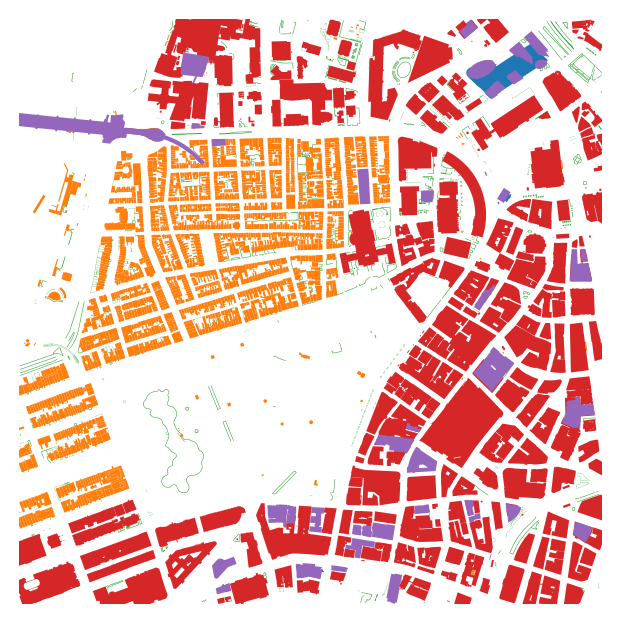} &
    \includegraphics[width=\sizefiggg\linewidth]{imgs/LOD/LOD_BOS_I_3.png}\\\hline
    \rotatebox{90}{\qquad \ \ \ 5} & \includegraphics[width=\sizefiggg\linewidth]{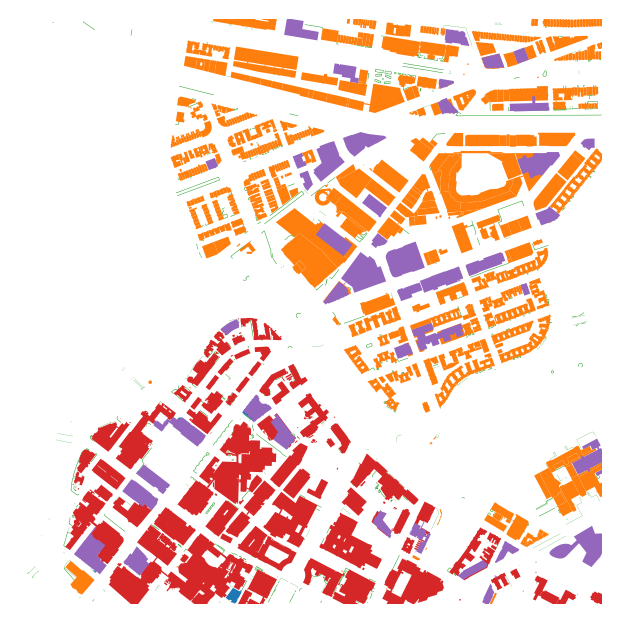} &
    \includegraphics[width=\sizefiggg\linewidth]{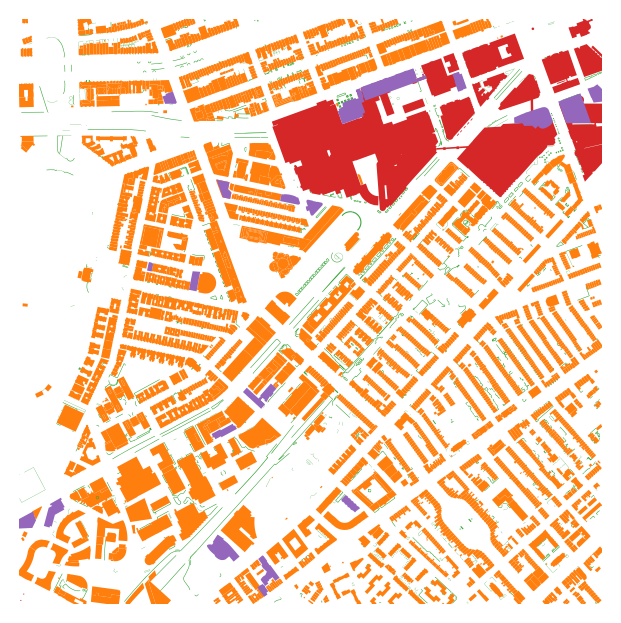} &
    \includegraphics[width=\sizefiggg\linewidth]{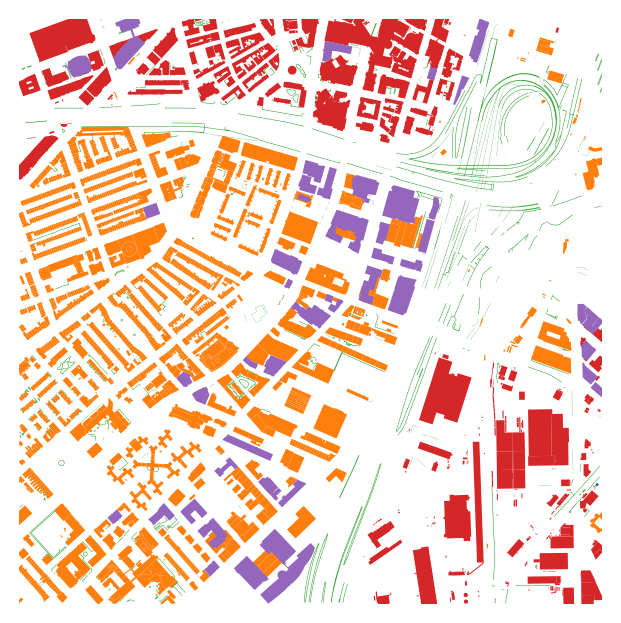} &
    \includegraphics[width=\sizefiggg\linewidth]{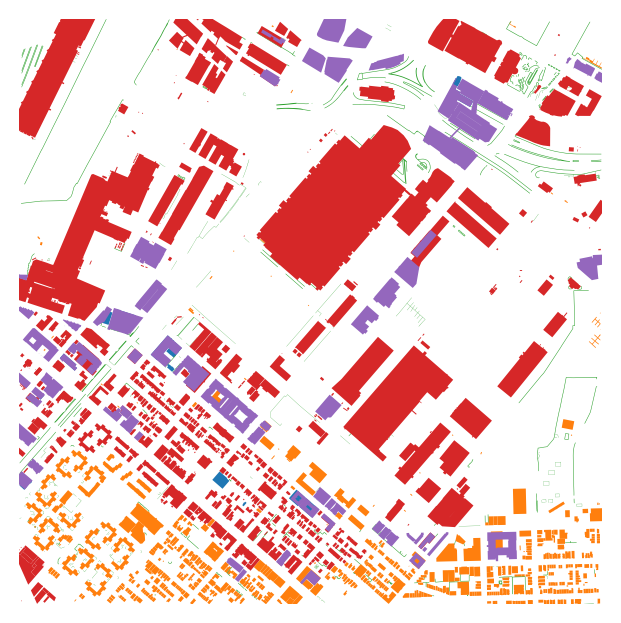}\\\hline
    \rotatebox{90}{\qquad \ \ \ 6} & \includegraphics[width=\sizefiggg\linewidth]{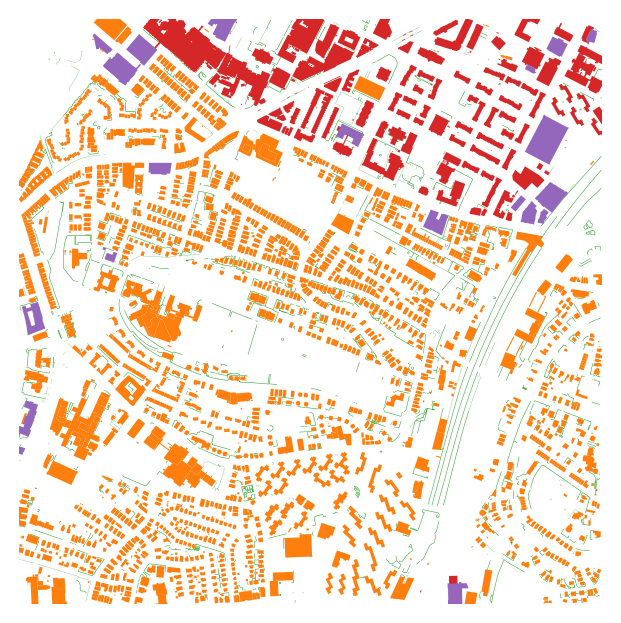} &
    \includegraphics[width=\sizefiggg\linewidth]{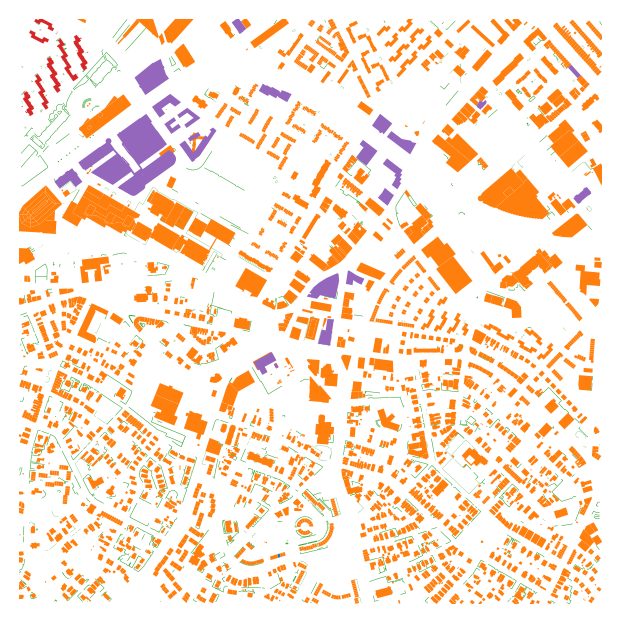} &
    \includegraphics[width=\sizefiggg\linewidth]{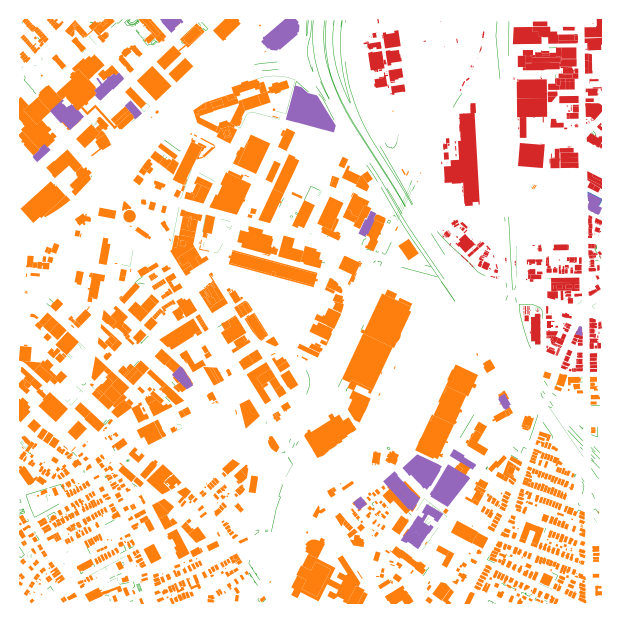} &
    \includegraphics[width=\sizefiggg\linewidth]{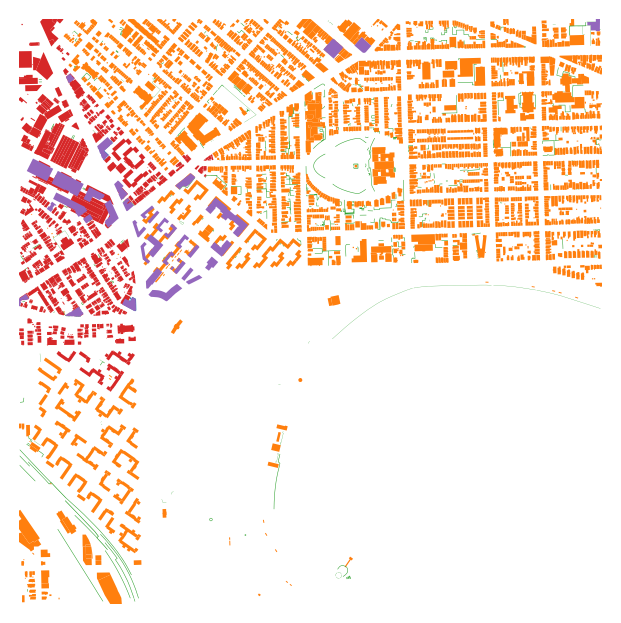}\\\hline
     \end{tabular}
    \caption{\gls{lod} of the building models in the central area of Boston.}
    \label{fig:lod_map}
\end{figure}

\glsreset{lod}
According to the citySchema format, each model has a different \gls{lod}.
The definition of the different \glspl{lod} of the standard are reported in~\cref{tab:lod}.
A map representing the \gls{lod} of the models in Downtown Boston is shown in~\cref{fig:lod_map}.

While the citySchema offers numerous advantages, the data format is not suitable for physics simulations or programmatic access. First, all the data has to be converted to \gls{iso} units, i.e., in meters. Second, the custom \gls{crs} makes it difficult to include data from other sources, e.g., the geolocated data. This limits the possibility of creating a digital twin with additional elements in the scenario, which is imperative for some use cases, e.g., in wireless communication network design. Finally, although the \gls{bpda} website and dataset offer numerous resources for visualization, catalog queries, and importing the models in commercial applications, the resources to access the data through programming languages such as Python are limited. 
Therefore, we decided to keep the same tile-based organization, while converting the models to a different \gls{crs} and format, and providing a georeferencing \gls{api}, described in \cref{sec:workflow}.

Specifically, the unit of the OBJ models was converted to meters and the meshes were centered in $[0,0]$, so that the vertex coordinates are all referred to the center of the model itself.
On one hand, this streamlines the visualization and the processing of the single meshes.
On the other, it moves the burden of defining the \gls{crs} to the user, who can thus define a custom one 
more easily.
To reverse the translation and geolocate the meshes, we include the geographic reference for each model in the model catalog of each tile (BOS\_<L>\_<N>.geojson), along with the information for the tile itself (BOS\_<L>\_<N>\_tileinfo.geojson).

Furthermore, we convert the OBJ models to PLY binary files, which is another popular, standardized data format that is supported (and encouraged~\cite{jakob2022mitsuba3}) by most of the open-source 3D processing and rendering libraries and tools, e.g., Blender~\cite{blender}, open3D~\cite{zhou2018open3d}, and Mitsuba~\cite{jakob2022mitsuba3}.
The resulting directory and file tree are reported in \cref{fig:dataset_dirs}, and publicly available at~\cite{bostontwin_data}.

\subsection{The \texttt{BostonAntennas} Class}
\label{ssec:boston_antennas}
The Boston \gls{gis} department released also the geolocated data for the requests to install small cell antennas/\gls{das} that were approved by the City of Boston.
The data was released in two datasets, ``\gls{das}/Small Cell Approved Locations''\cite{das_data} and ``Wireless Antenna Installation Requests Approved''\cite{antennas_data} for the requests approved before and after the 01/01/2017, respectively.
The \gls{crs} of both datasets is \gls{wgs84}.
We merged them into a single dataset, uniforming the column names and information (antennas.geojson in \cref{fig:dataset_dirs}).

Differently from the \texttt{BostonModel} class, the antenna data is 2D and contains no information on the elevation of the antennas.
However, in most cases, the type of mount where the antenna was installed (\textit{New\_Pole\_Type}) identifies a possible set of heights~\cite{poles}.

\subsection{The \texttt{BostonTwin} Class}
The digital twin is implemented through the \texttt{BostonTwin} class.
The class contains a (\texttt{BostonModel}, \texttt{BostonAntennas}) pair to access the information on the city-scale model.
An instance of the \texttt{BostonTwin} represents a specific \textit{scene}, and several utility functions to access it.

A scene is defined by its geographic boundaries (<scene\_name>\_ tileinfo.geojson in \cref{fig:dataset_dirs}) and includes all the digital twin elements within the corresponding area, i.e., the 3D model of the environment and the location of the base stations in the area.
Specifically, a scene can be accessed through the corresponding model catalog, or through the Mitsuba scene files (<scene\_name>.geojson and <scene\_name>.xml in \cref{fig:dataset_dirs}).
The Mitsuba files contain some basic settings for the 3D rendering of the scene, the reference to all the models in a scene, and the information to position the meshes within the scene local \gls{crs}.
The origin of the local coordinate system, in meters, is at the center of the scene.
The Mitsuba scene, when composed of PLY meshes, can be directly imported into Sionna, where an interactive \gls{api} for visualization and rendering is available, as explained in \cref{sec:workflow}.
As mentioned in \cref{ssec:BostonModel}, we keep the tiling organization and defined a scene for each tile, keeping the corresponding names (BOS\_<L>\_<N>).
However, the user can define custom scenes either by adding the corresponding files to the data directory or by using the dedicated \gls{api}, as explained in \cref{fig:workflow}.

Finally, the material of the surfaces that interact with the \gls{em} signal has a significant impact on its propagation.
In Sionna, a set of \gls{itu} materials with the corresponding \gls{em} properties~\cite{itu-p-2040} is available for the 3D meshes.
Each mesh is associated with a material.
In the predefined scenes of \projName, the materials are assigned based on the type of model reported in the Model Catalog (\textit{ITU brick} for "Wall", \textit{ITU concrete} for "Building", \textit{ITU medium-dry ground} for the ground).
The user can easily define new materials or change and customize the current ones through the Sionna \gls{api} or in the Mitsuba scene files, leveraging the georeferencing and descriptive information provided in the Model Catalog.
As the \gls{lod} of the \gls{bpda} data increases in future releases, we plan to support meshes with components of different materials.

\section{The BostonTwin Pipeline}
\label{sec:workflow}
\begin{figure*}[t]
    \centering
    \includegraphics[width=\textwidth]{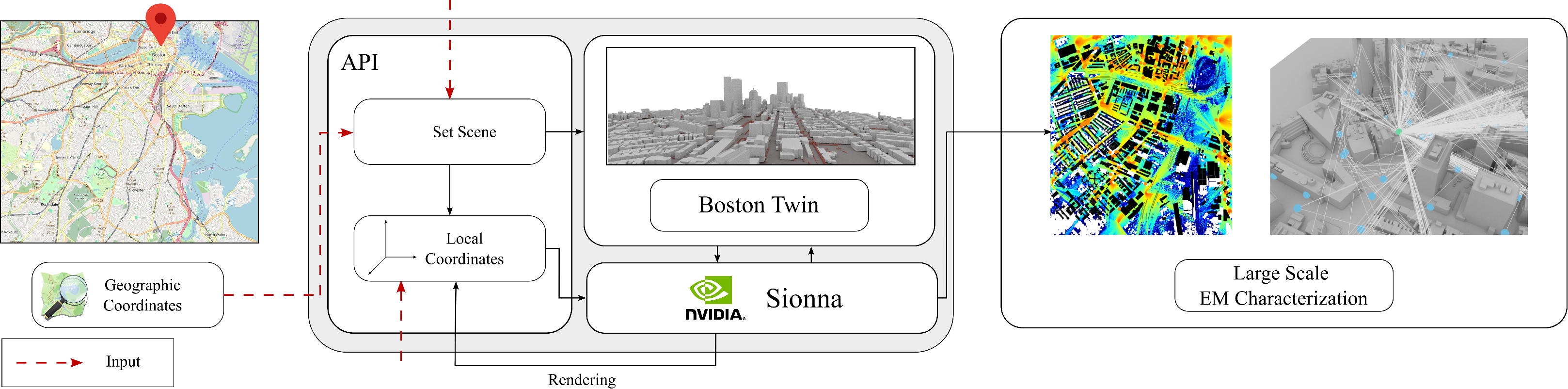}
    \caption{The workflow for the Boston Twin.}
    \label{fig:workflow}
\end{figure*}

In this section, we describe the workflow reported in \cref{fig:workflow} and the \gls{api} that we provide to interact with the \projName framework.

\paragraph{Scene Loading}
The user can load one of the predefined scenes (tiles) by specifying the tile name according to the convention presented in \cref{ssec:BostonModel} (top red arrow in \cref{fig:workflow}).
When doing so, the scene is loaded into Sionna, and the base station database is filtered to extract only the base stations in the tile.
Alternatively, the user can specify a geographical location (left red arrow in \cref{fig:workflow}), either through a (Longitude, Latitude) tuple (default \gls{crs}: \gls{wgs84}/EPSG:4326), or through a GeoPandas GeoDataFrame~\cite{joris_van_den_bossche_2024_10601680} with a different \gls{crs}, and a radius.
Then, the elements within the radius from the specified point are added to a new scene.

Thanks to the georeferencing capabilities of the \gls{api}, \projName can be easily integrated with data from any kind of sources, e.g., from \gls{sumo}~\cite{SUMO2018}, a traffic simulator that can provide realistic traffic patterns to be injected into the digital twin.
Furthermore, the compatibility with the citySchema format makes it easy to integrate the 3D data from other datasets.

\paragraph{Radio Devices Deployment}
As reported in \cref{ssec:boston_antennas}, the digital twin contains the information on the location of the approved base stations within the City of Boston, that allows the deployment of the radio devices at realistic locations.
Once the scene has been loaded, a list of base stations within the corresponding area is available.
The user can then select which ones to deploy in Sionna, giving each a role (either as a Sionna Transmitter or Receiver).
Furthermore, the user can add devices at custom locations, that can be specified as geographic or local coordinates, and significantly facilitates the deployment of, e.g., users.

\paragraph{\gls{rf} Characterization through Ray Tracing}
As reported in \cref{fig:workflow}, \projName relies on the Sionna ray tracer to simulate the \gls{em} wave propagation within the digital twin.
The ray tracer is based on Mitsuba 3 and can parallelize the computation on the GPU, thus achieving unprecedented results in terms of speed and scalability (for further details, we refer the readers to~\cite{hoydis2023sionnart}).
Thanks to these characteristics, and to the possibility of rendering the digital twin through Mitsuba and interacting with it through the Sionna \gls{api}, we identified it as the most promising open-source candidate for our work.
As showcased in \cref{sec:results}, the \gls{rt} allows to obtain detailed coverage maps for large scenarios with a runtime in the order of seconds, even with hundreds of thousands of triangles (\cref{tab:tiles_info}).
Furthermore, the point-to-point channel between each pair of devices can be computed with an arbitrary number of reflections, and with or without diffracted and scattered rays.

\paragraph{3D Model Simplification}
Finally, \projName offers some utility functions that can further enhance the usability and the interaction with the digital twin.
Large-scale \gls{em} simulations can require a huge amount of memory, which depends also on the number of triangles~\cite{lecci2021accuracy}.
Thus, the 3D meshes of the building models of \projName can be simplified, reducing the number of triangles to cut both the hardware requirements and the simulation runtime.
To do so, we leverage the Open3D library, that offers several methods to simplify the meshes.
The simplified version of the 3D data can then be imported, loaded, and used through the same pipeline as the original one.
However, the reduction in the number of vertices should be used carefully, as decreasing the precision can lead to inconsistent geometries, besides degrading the accuracy of the \gls{em} characterization.

We provide a detailed documentation and guide on the BostonTwin \glspl{api} in the open-source code repository~\cite{bostontwin_repo}.

\section{Digital Twin for Immersive Communications}
\label{sec:results}

\begin{table}[t]
    \centering
        \caption{Network and ray-tracing parameters used to obtain the coverage maps.}
    \label{tab:network_params}
    \begin{tabular}{|c|c|c|}
        \hline
         & \textbf{Parameter} & \textbf{Value} \\\hline
          \parbox[c]{2mm}{\multirow{6}{*}{\rotatebox[origin=c]{90}{\textbf{Network}}}}
         & Carrier Freq. & 12.7 GHz \\\cline{2-3}
         & Bandwidth & 400 MHz \\\cline{2-3}
         & $\#$ of Sectors & 3\\\cline{2-3}
         & Sector Width & $\ang{120}$\\\cline{2-3}
         & Base Station Height & 10~m\\\cline{2-3}
         & Ant. Array & $4\times4$\\\cline{2-3}
         & Ant. Pattern & tr38901~\cite{hoydis2023sionnart, 3gpp.38.901}\\\hline
          \parbox[c]{2mm}{\multirow{6}{*}{\rotatebox[origin=c]{90}{\textbf{Ray Tracer}}}} & Reflection & \cmark \\\cline{2-3}
         & Refl. Order & 3 \\\cline{2-3}
         & Diffraction & \cmark \\\cline{2-3}
         & Scattering & \xmark \\\cline{2-3}
         & Launch Rays & $10^6$ \\\cline{2-3}
         & Sampling & Fibonacci \\\hline
          \parbox[c]{2mm}{\multirow{2}{*}{\rotatebox[origin=c]{90}{\textbf{Req.}}}} & \gls{xr} & 30 Mbps\cite{3gpp.38.838, gapeyenko2023standardization} \\\cline{2-3}
         & \gls{v2x} & 700 Mbps \cite{3gpp22186}\\\hline
    \end{tabular}

\end{table}

\renewcommand{\sizefiggg}{0.24}
\begin{figure*}[t]
\centering
\begin{subfigure}[b]{.6\columnwidth}
\setlength\tabcolsep{0pt}
\renewcommand{\arraystretch}{0}
\centering
\begin{tabular}{c|c|c|c|c|}
  \multicolumn{5}{c}{
\begin{tikzpicture}

\definecolor{darkgray176}{RGB}{176,176,176}
\definecolor{darkorange}{RGB}{255,140,0}

\begin{axis}[
colorbar,
colorbar style={ylabel={}},
colormap={mymap}{[1pt]
  rgb(0pt)=(0,0,0.5);
  rgb(22pt)=(0,0,1);
  rgb(25pt)=(0,0,1);
  rgb(68pt)=(0,0.86,1);
  rgb(70pt)=(0,0.9,0.967741935483871);
  rgb(75pt)=(0.0806451612903226,1,0.887096774193548);
  rgb(128pt)=(0.935483870967742,1,0.0322580645161291);
  rgb(130pt)=(0.967741935483871,0.962962962962963,0);
  rgb(132pt)=(1,0.925925925925926,0);
  rgb(178pt)=(1,0.0740740740740741,0);
  rgb(182pt)=(0.909090909090909,0,0);
  rgb(200pt)=(0.5,0,0)
},
hide x axis,
hide y axis,
point meta max=20,
point meta min=-10,
colorbar horizontal,
colorbar style={
    at={(0,0)},               
    anchor=south,    
    width=4.5cm,
    xlabel={SNR [dB]},
    xlabel style={
        at={(0.5,3)}
    }
},
colorbar/width=2.5mm,
]
\end{axis}

\end{tikzpicture}}\\
 & F & G & H & I\\\hline
\rotatebox{90}{\quad \ \ \ 3} & & &
\includegraphics[width=\sizefiggg\linewidth]{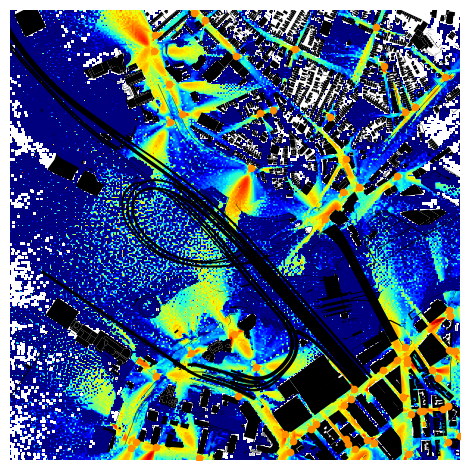} &
\includegraphics[width=\sizefiggg\linewidth]{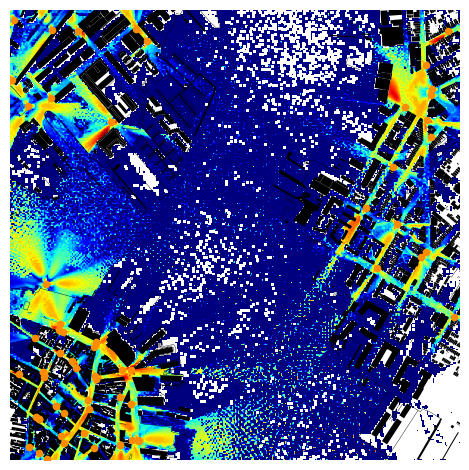}\\\hline
\rotatebox{90}{\quad \ \ \ 4} & \includegraphics[width=\sizefiggg\linewidth]{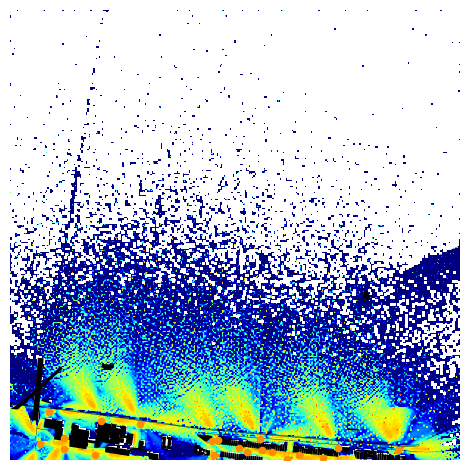} &
\includegraphics[width=\sizefiggg\linewidth]{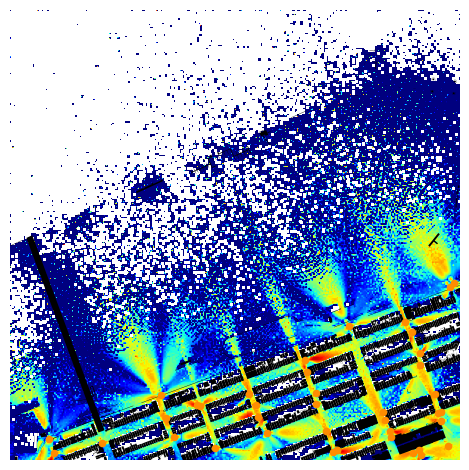} &
\includegraphics[width=\sizefiggg\linewidth]{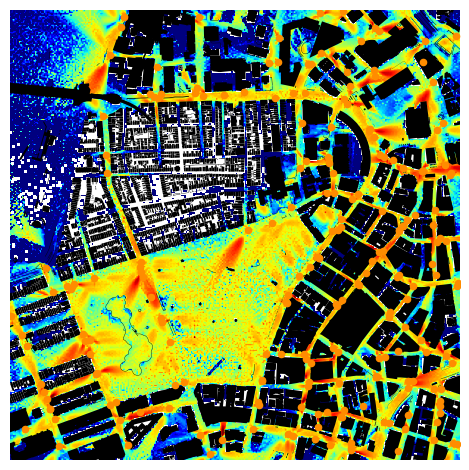} &
\includegraphics[width=\sizefiggg\linewidth]{imgs/cms_SNR/BOS_I_3.png}\\\hline
\rotatebox{90}{\quad \ \ \ 5} & \includegraphics[width=\sizefiggg\linewidth]{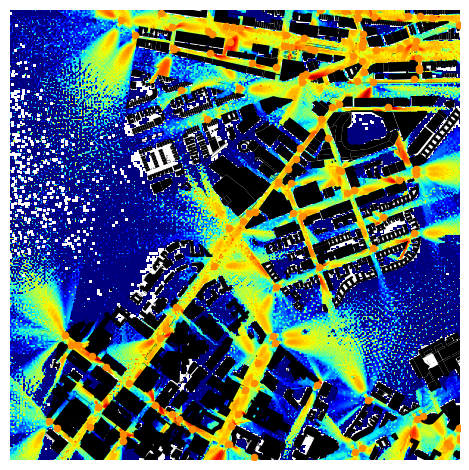} &
\includegraphics[width=\sizefiggg\linewidth]{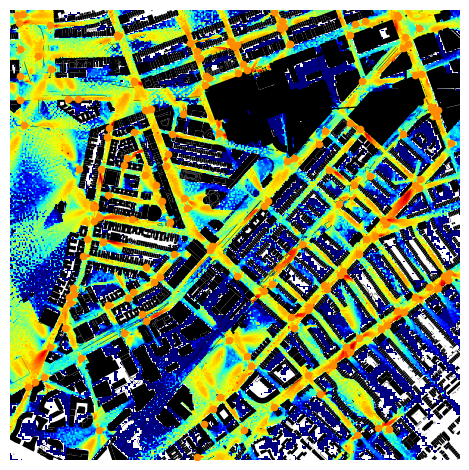} &
\includegraphics[width=\sizefiggg\linewidth]{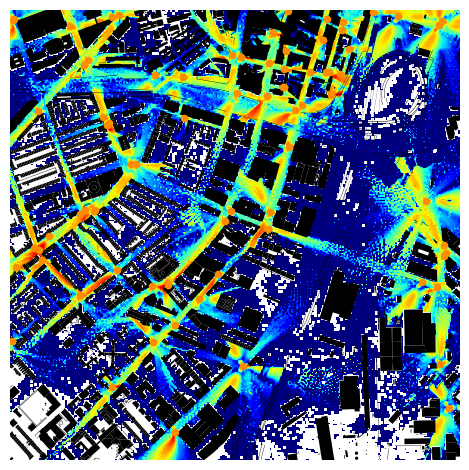} &
\includegraphics[width=\sizefiggg\linewidth]{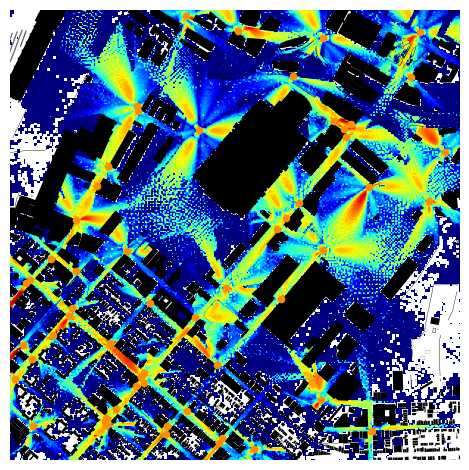}\\\hline
\rotatebox{90}{\quad \ \ \ 6} & \includegraphics[width=\sizefiggg\linewidth]{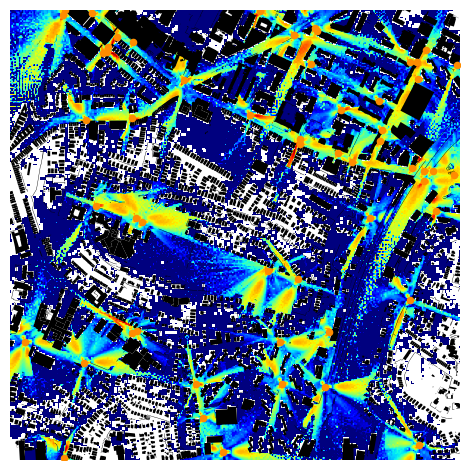} &
\includegraphics[width=\sizefiggg\linewidth]{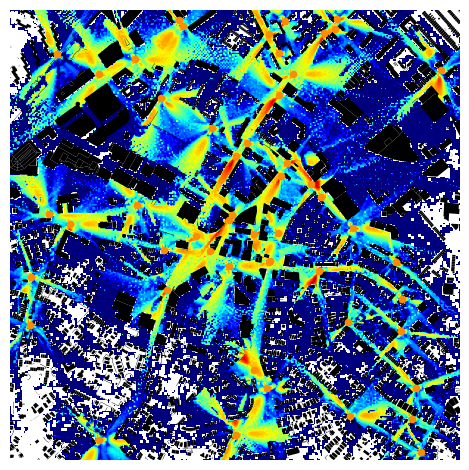} &
\includegraphics[width=\sizefiggg\linewidth]{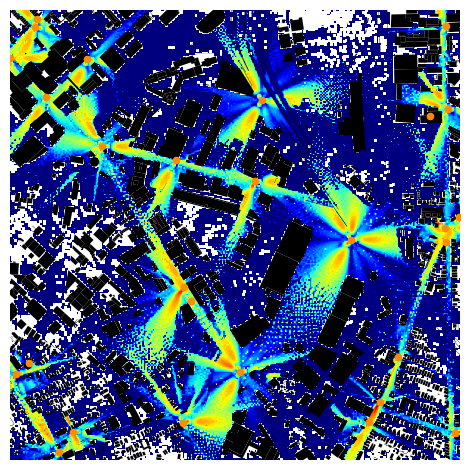} &
\includegraphics[width=\sizefiggg\linewidth]{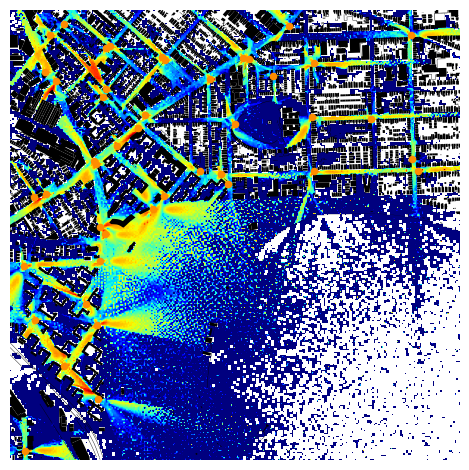}\\\hline
 \end{tabular}
\caption{}
\label{fig:cms_snr}
\end{subfigure}
\hfill
\begin{subfigure}[b]{.6\columnwidth}
    \setlength\tabcolsep{0pt}
\renewcommand{\arraystretch}{0}
\centering
\begin{tabular}{c|c|c|c|c|}
 & F & G & H & I\\\hline
\rotatebox{90}{\quad \ \ \ 3} & & &
\includegraphics[width=\sizefiggg\linewidth]{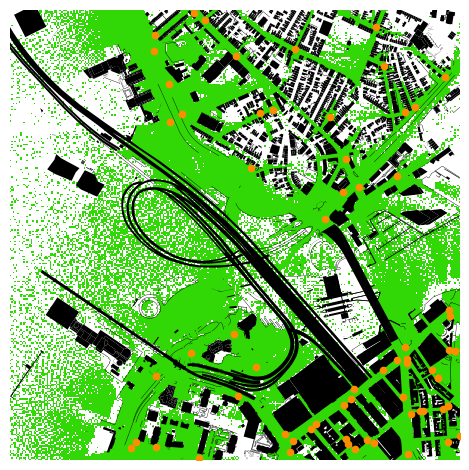} &
\includegraphics[width=\sizefiggg\linewidth]{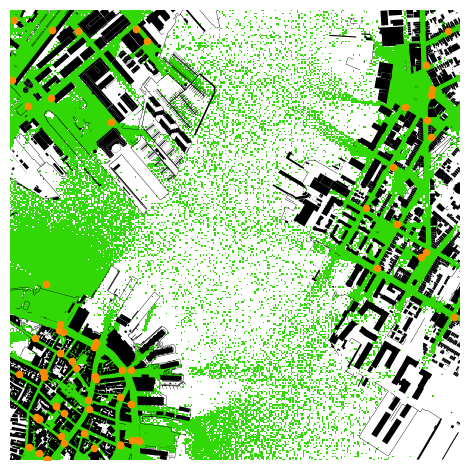}\\\hline
\rotatebox{90}{\quad \ \ \ 4} & \includegraphics[width=\sizefiggg\linewidth]{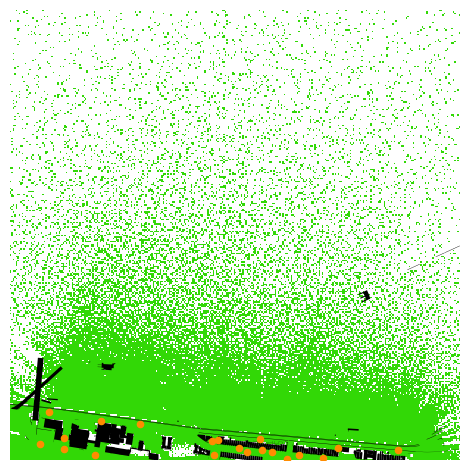} &
\includegraphics[width=\sizefiggg\linewidth]{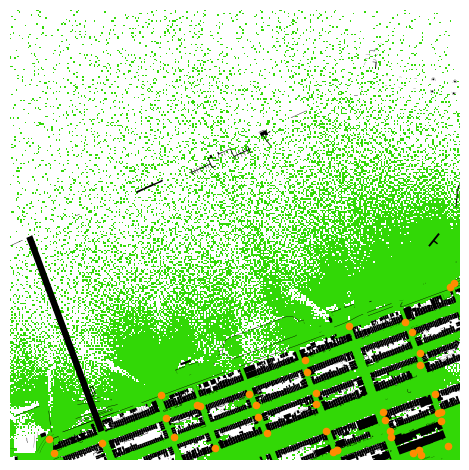} &
\includegraphics[width=\sizefiggg\linewidth]{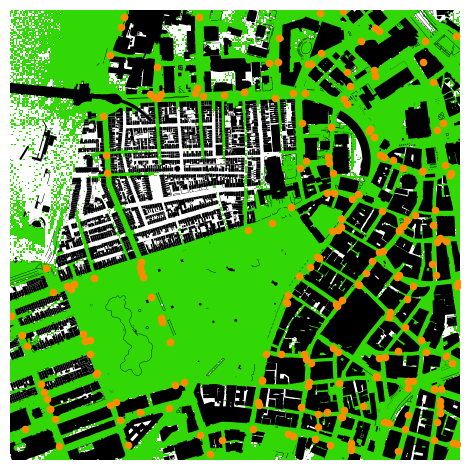} &
\includegraphics[width=\sizefiggg\linewidth]{imgs/cms_30Mbps/BOS_I_3.png}\\\hline
\rotatebox{90}{\quad \ \ \ 5} & \includegraphics[width=\sizefiggg\linewidth]{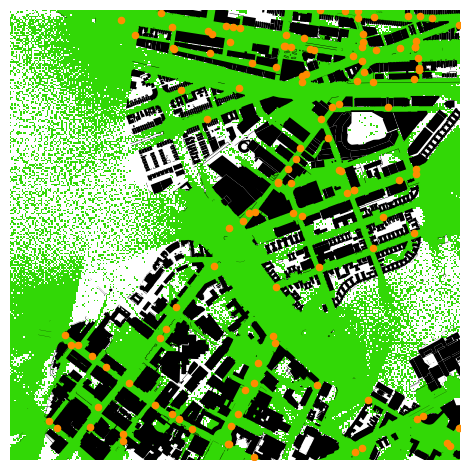} &
\includegraphics[width=\sizefiggg\linewidth]{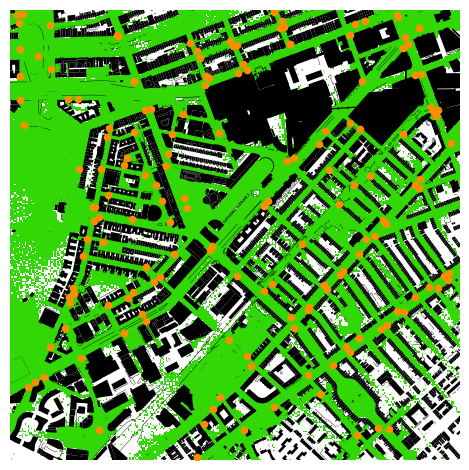} &
\includegraphics[width=\sizefiggg\linewidth]{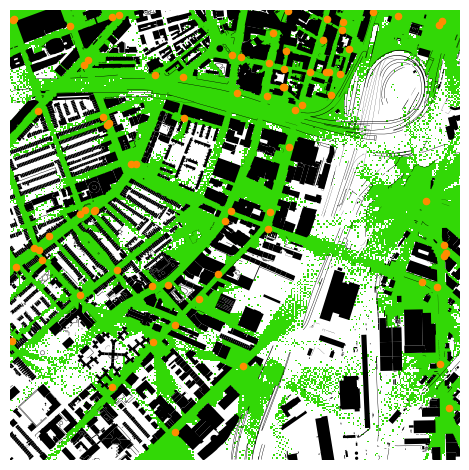} &
\includegraphics[width=\sizefiggg\linewidth]{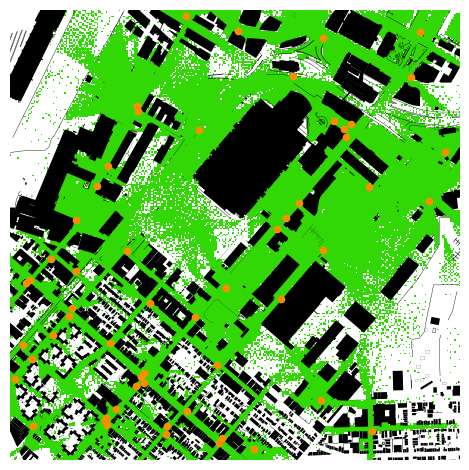}\\\hline
\rotatebox{90}{\quad \ \ \ 6} & \includegraphics[width=\sizefiggg\linewidth]{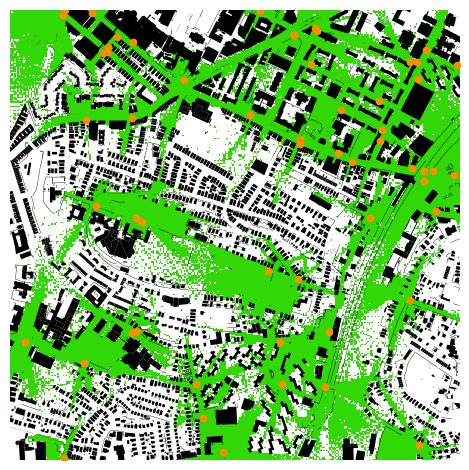} &
\includegraphics[width=\sizefiggg\linewidth]{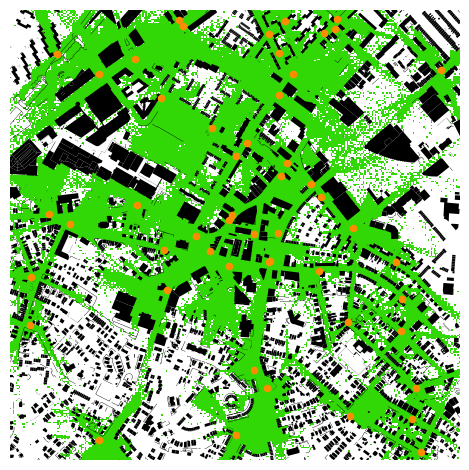} &
\includegraphics[width=\sizefiggg\linewidth]{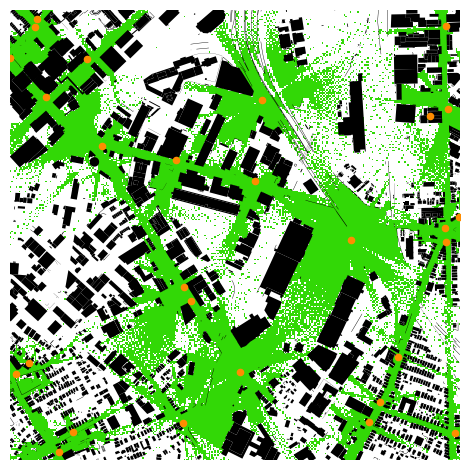} &
\includegraphics[width=\sizefiggg\linewidth]{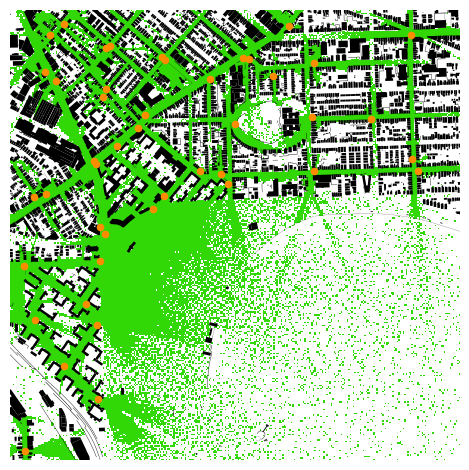}\\\hline
 \end{tabular}
\caption{}
\label{fig:cms_xr}
\end{subfigure}
\hfill
\begin{subfigure}[b]{.6\columnwidth}
    \setlength\tabcolsep{0pt}
\renewcommand{\arraystretch}{0}
\centering
\begin{tabular}{c|c|c|c|c|}
 & F & G & H & I\\\hline
\rotatebox{90}{\quad \ \ \ 3} & & &
\includegraphics[width=\sizefiggg\linewidth]{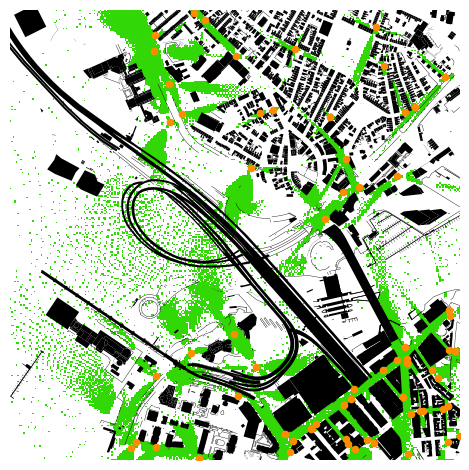} &
\includegraphics[width=\sizefiggg\linewidth]{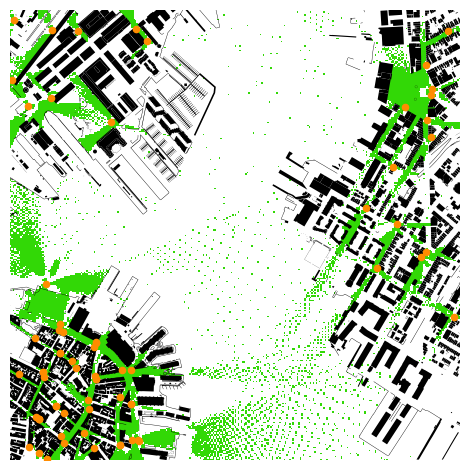}\\\hline
\rotatebox{90}{\quad \ \ \ 4} & \includegraphics[width=\sizefiggg\linewidth]{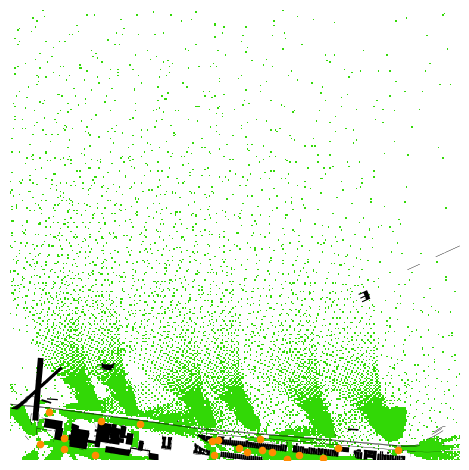} &
\includegraphics[width=\sizefiggg\linewidth]{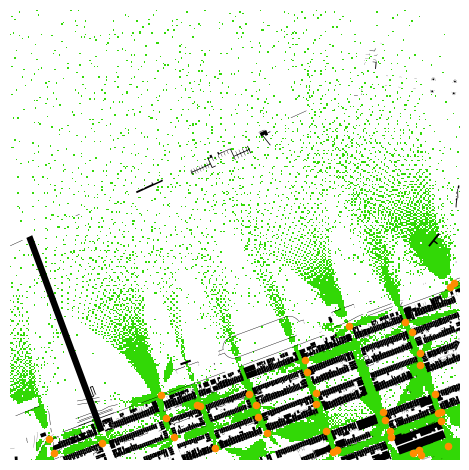} &
\includegraphics[width=\sizefiggg\linewidth]{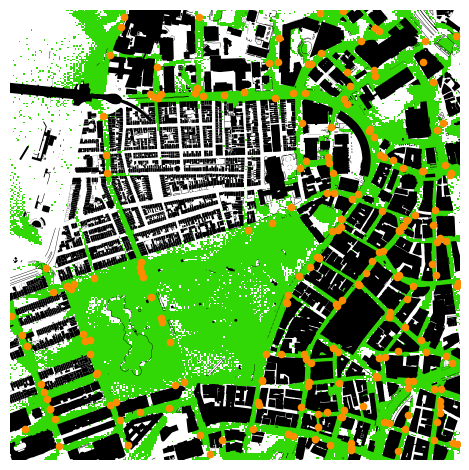} &
\includegraphics[width=\sizefiggg\linewidth]{imgs/cms_700Mbps/BOS_I_3.png}\\\hline
\rotatebox{90}{\quad \ \ \ 5} & \includegraphics[width=\sizefiggg\linewidth]{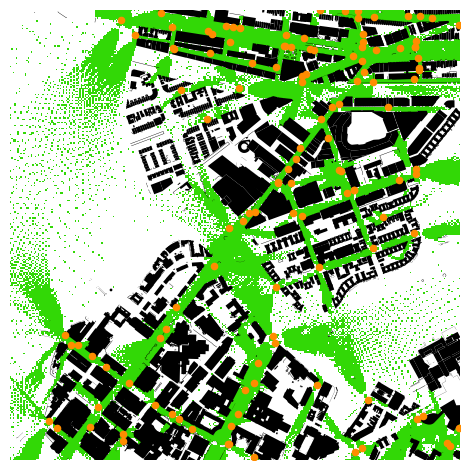} &
\includegraphics[width=\sizefiggg\linewidth]{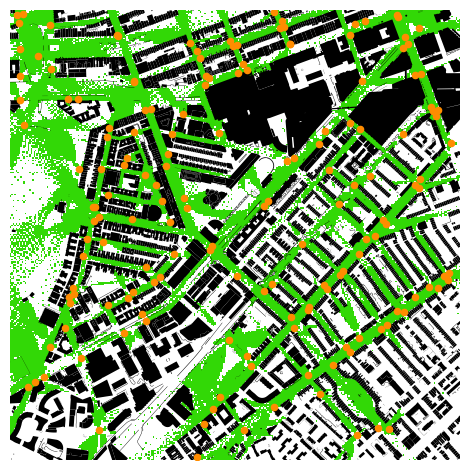} &
\includegraphics[width=\sizefiggg\linewidth]{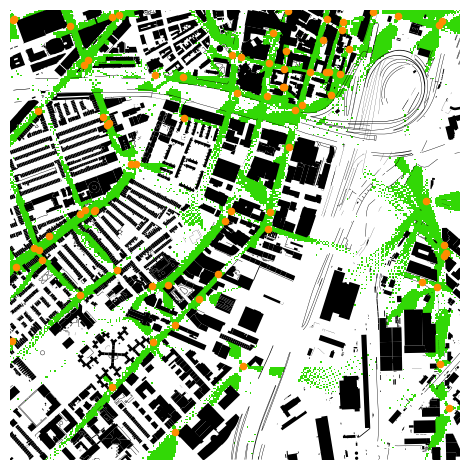} &
\includegraphics[width=\sizefiggg\linewidth]{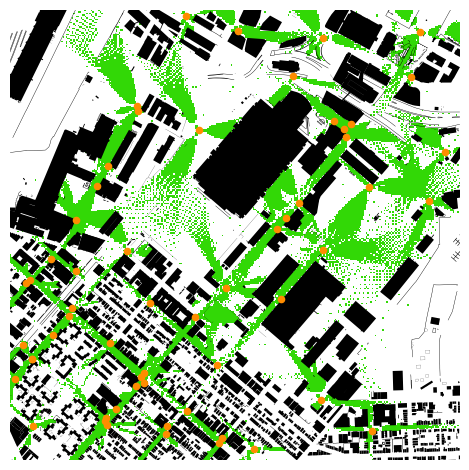}\\\hline
\rotatebox{90}{\quad \ \ \ 6} & \includegraphics[width=\sizefiggg\linewidth]{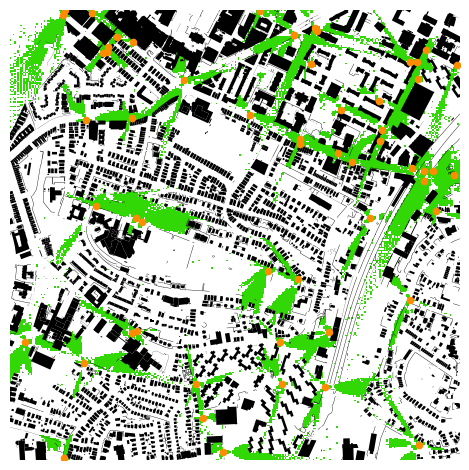} &
\includegraphics[width=\sizefiggg\linewidth]{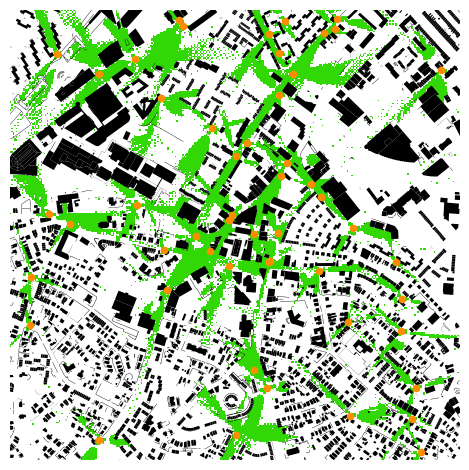} &
\includegraphics[width=\sizefiggg\linewidth]{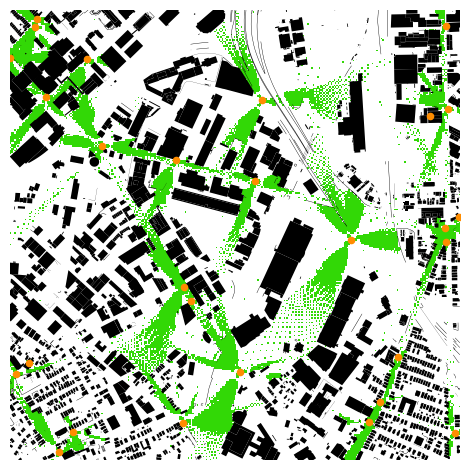} &
\includegraphics[width=\sizefiggg\linewidth]{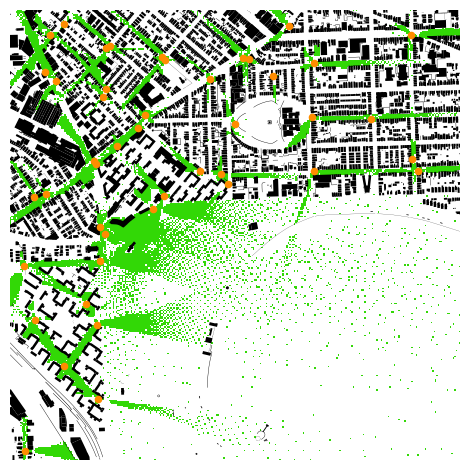}\\\hline
 \end{tabular}
\caption{}
\label{fig:cms_v2x}
\end{subfigure}
\caption{Coverage map for the central area of Boston in terms of \gls{snr} (a) and throughput requirements for \gls{xr} (b) and extended sensors in \gls{v2x} (c). In the latter two maps, the green areas indicate where the requirements are satisfied.}
\end{figure*}

We showcase the potential of \projName as a digital twin considering a simplified \gls{6g} scenario for immersive communications.
Specifically, we consider $12.7$~GHz as carrier frequency, one of the candidate frequencies in the novel FR-3 bands of interest for 6G systems~\cite{cui20236g,kang2023cellular}.
Particularly during the initial evaluation phase, when new frequency allocations are discussed, having reliable simulation tools is of the foremost importance, e.g., to assess the impact on existing stakeholders~\cite{testolina2023modeling}, and to characterize the propagation and the corresponding deployment design in different environments.
To do so, an accurate propagation model is necessary, but also a detailed representation of the propagation environment.

Thus, as FR-3 offers more bandwidth than the traditional FR-1 with better propagation characteristics than millimeter waves, we analyze whether the increased bandwidth can meet the requirements of two immersive communications technologies, i.e., \gls{xr} and Video Streaming in \gls{v2x}.
Following the workflow described in \cref{sec:workflow}, we selected 16 scenes (tiles BOS\_L\_N, L$\in \{\text{F}-\text{I}\}$, N$\in \{3-6\}$), representing a 83.6~km$^2$-wide area in the central part of Boston.
The complete network and ray tracer configuration parameters are reported in \cref{tab:network_params}.

We start by characterizing the \gls{snr} in each tile by modeling the propagation of \gls{em} waves using the Sionna ray tracer considering all the base stations in each tile and computing it over a $400$~MHz bandwidth $B$.
From \cref{fig:cms_snr} we can observe that thanks to the high density of the base stations, the downtown area (G\_4-5, H\_3-4, I\_3-4) is well serviced, with a peak \gls{snr} of more than 24~dB.
On the contrary, the southernmost areas have a lower base station density and thus a lower coverage.
The \gls{snr} where the signal can propagate in free space, i.e., without the building obstruction, is visible in the open areas such as the Charles River banks (e.g., F\_4, G\_4) or the Boston Common Park (central area in tile H\_4). 

Then, we compute the achievable capacity $C$ using the Shannon capacity equation $C = B \log_2(1+\textrm{SNR})$.
We consider the throughput requirement for two immersive communications use cases, i.e., 30 Mbps for \gls{xr}~\cite{3gpp.38.838,gapeyenko2023standardization,chiariotti2023temporal} and 700 Mbps for a use case defined in the 3GPP technical specification~\cite{3gpp22186}, i.e., ``video sharing between UEs supporting \gls{v2x} application'', and filter out the locations where the capacity does not meet the requirements.
\cref{fig:cms_xr} shows that the current base station can deliver enough throughput to support the datarate of the XR application in large areas of the city.
On the contrary, the capacity by the \gls{v2x} use case is met only in the immediate surroundings of the base stations, even when using the frequencies in FR-3, calling for a denser deployment, higher transmit power or even larger bandwidth.

Using \projName, it is thus possible to analyze the existing scenarios, plan and test further network deployment, study its dimensioning, and use it as a safe, yet realistic playground for innovative orchestration solutions on an unprecedented scale.

\section{Conclusions}
\label{sec:conclusions}

In this paper, we introduced a city-scale twinning framework that simplifies ray-tracing studies for next-generation wireless networks and multimedia applications. Starting from open data from the city of Boston, we have created a dataset that can be easily parsed, manipulated, and imported in popular \gls{rf} ray tracing frameworks. In the paper, we first described the dataset and code that was developed to interact with it. Then, we introduced a workflow that takes the 3D models, combines them with custom geolocated points of interest, and feeds this information to the NVIDIA Sionna ray tracer. In the last part of the paper, we focused on a use case that showcases the flexibility and scale of \projName, using \gls{rf} ray tracing in the accurate digital representation of the city of Boston to profile whether different types of next-generation multimedia applications can be supported by the wireless network in new frequency bands of interest for 6G. 

The data is publicly available at~\cite{bostontwin_data}, and the code is open source under the MIT license and available at~\cite{bostontwin_repo}. As future work, we will leverage the \projName framework to create realistic \gls{rt}-based scenarios for the city of Boston in Colosseum, enabling full-stack twinning studies in cellular and wireless systems, and compare with real-world measurements for validation. In addition, we will embed a more advanced terrain model, including elevation, and integrate with mobility simulators such as \gls{sumo}.


\begin{acks}

This work was partially supported by the National Telecommunications and Information Administration (NTIA)'s Public Wireless Supply Chain Innovation Fund (PWSCIF) under Award No. 25-60-IF011.

\end{acks}

\bibliographystyle{ACM-Reference-Format}
\bibliography{bibliography}

\end{document}